\documentclass[review]{elsarticle}
\usepackage{amsmath}
\usepackage{lineno}
\modulolinenumbers[5]
\usepackage{subfigure}
\usepackage{tikz}

\usepackage{float}

\usepackage{graphicx}
\usepackage[]{hyperref}
\hypersetup{hidelinks}
\pdfoutput=1
\begin{document}

\begin{frontmatter}

\title{A Novel Finite Difference Method for Euler Equations in 2D Unstructured Meshes}

\author[mymainaddress]{
   Meiyuan Zhen
}
\author[mymainaddress]{
   Kun Qu\corref{mycorrespondingauthor}
}
\author[mymainaddress]{
   Jinsheng Cai
}

% \fntext[myfootnote]{Since 1880.}
\cortext[mycorrespondingauthor]{Corresponding author}

\address[mymainaddress]{School of Aeronautics, Northwestern Polytechnical University, Xi'an 710072, Shaanxi, China.}
% \address[mysecondaryaddress]{360 Park Avenue South, New York}
% \maketitle

\begin{abstract}
Finite difference method was extended to unstructured meshes to solve Euler equations. The spatial discretization is made of two steps. 
First, numerical fluxes are computed at the middle point of each edge with high order accuracy. In this step, a one-dimensional curvilinear stencil is assembled for each edge to perform one-dimensional fast non-uniform WENO interpolation derived in this paper, which is much easier and faster than multi-dimensional interpolation. The second step is to compute the divergence of fluxes at each vertex from the fluxes at nearby edges and vertices by least square approximation of multi-dimensional polynomials. The order of the WENO interpolation in the first step and the degree of the polynomial in the second step determined the order of accuracy of the spatial scheme.  After that, explicit RungeKutta time discrete scheme is used to update conservative variables. Several canonical numerical cases were solved to test the accuracy, performance and the capability of shock capturing of the developed method.
\end{abstract}


\end{frontmatter}

% \linenumbers

%%%% Start %%%%%%
\section{Introduction} \label{sec_intro}
In decades, the research of high order methods is one of the most important work in CFD,
 with a great number of high order methods being proposed.
Among them, finite difference method (FDM) \cite{jiang_efficient_1996, henrick_mapped_2005,borges_improved_2008,hu_adaptive_2010}, discontinuous Galerkin (DG) \cite{reed1973triangular, cockburn1989tvb, cockburn2001runge}, flux reconstruction (FR) \cite{huynh_flux_2007,witherden_pyfr_2014}, spectral difference method (SDM) \cite{liu2006spectral,wang2007spectral}, as well as high order finite volume method (FVM), \cite{barth_higher_1990, hu_weighted_1999, wang_compact_2016} are studied most intensively.
% FD
Today, high order finite difference schemes are so matured that they are able to resolve shear layers, vortices and capture shock waves and material interfaces in rectangular or curve-linear grids \cite{wong_high-order_2017, su2020numerical}.
Besides, the performance of FDM is very high since only one-dimensional operations are necessary.
High order of accuracy in FDM can be easily implemented. Fifth or higher order schemes are widely used in the simulation with FDM.
Thus high order FDM are adopted with DNS/LES in a wide range of researches such as flow instability and multi-phase flows.
Although highly developed, the shortcoming of FDM is obvious that generating structural grids is very difficult for complex geometries. This hinders its applications in engineering problems.
% DG/FR
Thus, high order methods on unstructured meshes, such as DG, FR and SD, attract more scientists. Different from FDM, these methods rely on elements with multiple degrees of freedom. In each element, polynomials serve as shape functions, which naturally leads to the difficulty of capturing strong shock waves.
Although some measures such as adding artificial viscosity and slope limiters or combined with FVM are proposed to overcome this problem, which increase the complexity of the method and therefore still needs to be improved.

At the same time, some researchers tried to develop generalized finite difference method (GDM) in the context of meshless methods \cite{lacaze2009large, li_high_2020}.
In meshless methods, on a virtual edge between two closed points, numerical fluxes are calculated with a Riemann solver and utilized to discretize the divergence of the flux on each node.
The weights of computing divergence are determined from some kind of local kernel function based on distance.
This algorithm can capture shock waves since it adopted Riemann solvers.
Usually, the closest points are used and only the 2nd order accuracy.
In the work of Li and Ren \cite{li_high_2020}, more points are introduced into each stencil to achieve higher order accuracy.

In this paper, the authors developed a FDM on unstructured meshes.
In this method, similar to Ren's work, the divergence of flux in the governing equations describing a conservative law is calculated on each vertex directly from the nearby vertices and edges by means of least square, without any kind of real or virtual control volume and integrals on faces or volumes.
The two differences are: 1) a stencil is constructed based the topology of the mesh and able to adapt to the local non-isentropic characteristics. It also includes more edges thus making the stencil more compact; 2) the numerical flux on each edge is computed from a one dimensional curve stencil but not multi-dimensional stencil, making our method more efficient. To the authors' knowledge, it is the first attempt to extract one-dimensional interpolation stencil in the context of unstructured meshes.

The rest of this paper is organized as follows. Section \ref{sec2} describes several key components of the developed method, including flux divergence computing and numerical flux computing on one-dimensional stencils. Section \ref{sec3} gives several numerical test cases to verify the performance of the method. Conclusions and possible future work are presented in section \ref{sec4}.
\section{Numerical Method}
\label{sec2}
\subsection{Governing Equations} \label{sec2.1}
The two-dimensional Euler equation describes inviscid compressible flow, which is given by
\begin{equation}
  \frac{\partial \mathbf{w}}{\partial t}
+ \frac{\partial \mathbf{f}(\mathbf{w})}{\partial x}
+ \frac{\partial \mathbf{g}(\mathbf{w})}{\partial y} = 0
\end{equation}
where
$$
\mathbf{w} = \left[\begin{array}{c}\rho \\ \rho u \\ \rho v \\ \rho E \end{array}\right]
\quad
\mathbf{f}(\mathbf{w})
= \left[\begin{array}{c}\rho u \\ \rho uu +p \\ \rho u v \\ (\rho E+p) u \end{array}\right]
\quad
\mathbf{g}(\mathbf{w})
= \left[\begin{array}{c}\rho v \\ \rho vu  \\ \rho v v + p \\ (\rho E+p) v \end{array}\right]
$$
and $\rho$, $u$, $v$, $p$, $E$ are density, velocity in $x$, $y$ directions, pressure and total energy respectively, where $E=e+\frac{u^2+v^2}{2}$ and $p=(\gamma-1)\rho e$. $\gamma=1.4$ is used in this work. In the following, scalar $f$ and $g$ are used to denote a component in the flux vector $\mathbf{f}$ and  $\mathbf{g}$ respectively.

%section 2.2------------------------------
\subsection{Discretization of Flux Divergence}\label{sec2.2}
From the governing equations, it can be seen that the task of spatial discretization is to compute $\partial_x f + \partial_y g$ which is the divergence of fluxes.

%% Fig of the stencil based on node/edge flux
For structured grids, the flux divergence at a node is calculated along 2(in 2D cases) or 3(in 3D cases) grid lines with some FD schemes.
For instance, the first order derivative of the flux computed in WCNS (Weighted Compact Nonlinear Scheme)\cite{deng_developing_2000} is approximated with high order central difference from the numerical fluxes at nearby middle points or both middle points and nodes. Artificial viscosity is not used explicitly as numerical dissipation is already introduced when computing the numerical flux at middle points, making the flux function smooth enough to perform central FDM.

When it comes to unstructured meshes used herein, similar to the way that WCNS used, flux divergence is computed based on the numerical flux on each edge and analytical flux vector on each vertex, which is described in the following subsections.
%%
%%  Fig of FDM
%%
%section 2.2.1-------------------
\subsubsection{Least-Square Based Approximation of Flux Divergence}
Herein, the two flux functions $f$ and $g$ around a vertex $O$  are approximated by two dimensional polynomials
\begin{align}
  f(x,y) &=f_0 + a_1 x + a_2 y + a_3 x^2 + a_4 xy + a_5 y^2 + \cdots = f_0 + \mathbf{p}^T(x,y) \mathbf{a} \label{ef} \\
  g(x,y) &=g_0 + b_1 x + b_2 y + b_3 x^2 + b_4 xy + b_5 y^2 + \cdots = g_0 + \mathbf{p}^T(x,y) \mathbf{b}\label{eg}
\end{align}
where $f_0$ and $g_0$ are the flux components at the vertex $O$ which is regarded as the origin point $(0,0)$.
Vector
\begin{equation}
  \mathbf{p}(x,y) = \left[x,y,x^2,xy,y^2,x^3,x^2y,xy^2,y^3, \cdots \right]^T
\end{equation}
contains $M$ necessary terms of two-dimensional polynomials.
Eq.\eqref{ef} and Eq.\eqref{eg} should be satisfied in a domain near the vertex $O$. For a vertex $j$ nearby,
\begin{align}   
  \mathbf{p}^T(x_j,y_j) \mathbf{a} &= f_j -f_0 \\
  \mathbf{p}^T(x_j,y_j) \mathbf{b} &= g_j -g_0
\end{align}
For an edge $k$, only the projection of the flux vector $[f,g]^T$ at the middle-edge point along the direction of the edge is provided. Thus,
\begin{equation}
  n_{x,k} \mathbf{p}^T(x_k,y_k) \mathbf{a} + n_{y,k} \mathbf{p}^T(x_k,y_k) \mathbf{b}  =
   r_k - (n_{x,k}f_0 + n_{y,k} g_0)
\end{equation}
where $r_k$ is the flux projection at the middle point of edge $k$ and $[n_{x,k},n_{y,k}]^T$ represents the unit directional vector of edge $k$.
Picking enough vertices and edges nearby, an overdetermined linear system can be assembled as
\begin{equation}
\left[\begin{array}{cc}
\mathbf{p}^T(x_{j=1},y_{j=1})   &   O \\
O                       & \mathbf{p}^T(x_{j=1},y_{j=1}) \\
\mathbf{p}^T(x_{j=2},y_{j=2})   &   O \\
O                       & \mathbf{p}^T(x_{j=2},y_{j=2}) \\
\vdots  & \vdots  \\
\mathbf{p}^T(x_{j=J},y_{j=J})   &   O \\
O                       & \mathbf{p}^T(x_{j=J},y_{j=J})  \\
n_{x,k=1} \mathbf{p}^T(x_{k=1},y_{k=1})  &   n_{y,k=1} \mathbf{p}^T(x_{k=1},y_{k=1})  \\
n_{x,k=2} \mathbf{p}^T(x_{k=2},y_{k=2})  &   n_{y,k=2} \mathbf{p}^T(x_{k=2},y_{k=2})  \\
\vdots & \vdots \\
n_{x,k=K} \mathbf{p}^T(x_{k=K},y_{k=K})  &   n_{y,k=K} \mathbf{p}^T(x_{k=K},y_{k=K})
\end{array}\right]
\begin{bmatrix}
   \mathbf{a}\\
   \mathbf{b}
\end{bmatrix}
= \mathbf{R} 
\label{eLinearSystem}
\end{equation}
where
\begin{equation}
\mathbf{R}=
\left[\begin{array}{c}
f_{j=1} - f_0 \\
g_{j=1} - g_0 \\
f_{j=2} - f_0 \\
g_{j=2} - g_0 \\
\vdots \\
f_{j=J} - f_0 \\
g_{j=J} - g_0 \\
r_{k=1} - (n_{x,k=1}f_0 + n_{y,k=1} g_0) \\
r_{k=2} - (n_{x,k=2}f_0 + n_{y,k=2} g_0) \\
\vdots \\
r_{k=K} - (n_{x,k=K}f_0 + n_{y,k=K} g_0)
\end{array}\right]
\end{equation}
The linear system Eq.\eqref{eLinearSystem} can be simply put as
\begin{equation}
   \mathbf{A}
   \begin{bmatrix}
      \mathbf{a}\\
      \mathbf{b}
   \end{bmatrix}
 = \mathbf{R}   
\end{equation}
 which can be solved with normal matrix method or SVD, with later used herein,

\begin{equation}
   \begin{bmatrix}
      \mathbf{a}\\
      \mathbf{b}
   \end{bmatrix} =
\mathbf{A}^{\dagger} {\mathbf{R}}
\iffalse
\left[\begin{array}{c}
f_1 - f_0 \\
g_1 - g_0 \\
f_2 - f_0 \\
g_2 - g_0 \\
\vdots \\
f_J - f_0 \\
g_J - g_0 \\
r_1 - (n_{x,1}f_0 + n_{y,1} g_0) \\
r_2 - (n_{x,2}f_0 + n_{y,2} g_0) \\
\vdots \\
r_K - (n_{x,K}f_0 + n_{y,K} g_0)
\end{array}\right]
\fi
\end{equation}
where $\mathbf{A}^{\dagger}$ is the pesudo-inverse of $\mathbf{A}$.
It is noticed that by utilizing the polynomial approximation in Eq.\eqref{ef} and Eq.\eqref{eg}, the flux divergence at Vertex $O$ can expressed as weighted sum of $\mathbf{R}$ components,
\begin{equation}
\left. \frac{\partial f}{\partial x} \right|_{x=0,y=0} +
\left. \frac{\partial g}{\partial y} \right|_{x=0,y=0}
= a_1  +  b_2
= (\mathbf{r}^{\dagger}_{a_1}  +  \mathbf{r}^{\dagger}_{b_2})  {\mathbf{R}}
\end{equation}
where $\mathbf{r}^{\dagger}_{a_1}$ and $\mathbf{r}^{\dagger}_{b_2}$ are the row vectors in $\mathbf{A}^{\dagger}$ corresponding to $a_1$ and $b_2$ respectively.
The vector $\mathbf{r}^{\dagger}_{a_1}$ + $\mathbf{r}^{\dagger}_{b_2}$ is only related to the geometry of mesh and can be computed only once at the initial time for static meshes.

By building stencils with enough vertices and edges around a vertex, high order accuracy of the flux divergence can be achieved easily.

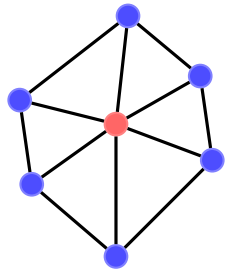
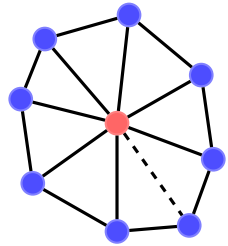
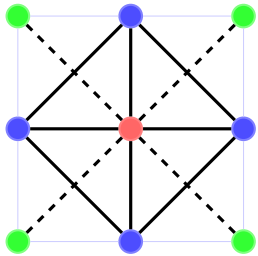
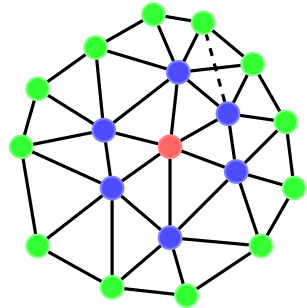
\begin{figure*}[hbtp]
  \centering
  \tikzstyle{nodeI}=[circle,draw=red!50, fill=red!60,thick, inner sep=0pt,minimum size=3mm]
  \tikzstyle{nodeJ}=[circle,draw=blue!50,fill=blue!70, thick, inner sep=0pt,minimum size=3mm]
  \tikzstyle{nodeK}=[circle,draw=green!50,fill=green!80, thick, inner sep=0pt,minimum size=3mm]
  \subfigure[A one-level stencil in a triangular mesh.]{
    \label{l1-tri}
    \centering
\begin{tikzpicture}[scale=1.6]
  \node[shape=circle,fill=gray] (I)  at (+0.0,+0.0) [nodeI] { };
  \node[shape=circle,fill=gray] (J1) at (+0.8,-0.3) [nodeJ] { };
  \node[shape=circle,fill=gray] (J2) at (+0.7,+0.4) [nodeJ] { };
  \node[shape=circle,fill=gray] (J3) at (+0.1,+0.9) [nodeJ] { };
  \node[shape=circle,fill=gray] (J4) at (-0.8,+0.2) [nodeJ] { };
  \node[shape=circle,fill=gray] (J5) at (-0.7,-0.5) [nodeJ] { };
  \node[shape=circle,fill=gray] (J6) at (-0.0,-1.1) [nodeJ] { };
  \draw [very thick] (I) -- (J1);
  \draw [very thick] (I) -- (J2);
  \draw [very thick] (I) -- (J3);
  \draw [very thick] (I) -- (J4);
  \draw [very thick] (I) -- (J5);
  \draw [very thick] (I) -- (J6);
  \draw [very thick] (J1) -- (J2);
  \draw [very thick] (J2) -- (J3);
  \draw [very thick] (J3) -- (J4);
  \draw [very thick] (J4) -- (J5);
  \draw [very thick] (J5) -- (J6);
  \draw [very thick] (J6) -- (J1);
  \end{tikzpicture}
 }
 \hspace{0.5in}
  \subfigure[A one-level stencil in a hybrid mesh.]{
    \label{l1-hyb}
    \centering
  \begin{tikzpicture}[scale=1.6]
    \node[shape=circle,fill=gray] (I)  at (+0.0,+0.0) [nodeI] { };
    \node[shape=circle,fill=gray] (J1) at (+0.8,-0.3) [nodeJ] { };
    \node[shape=circle,fill=gray] (J2) at (+0.7,+0.4) [nodeJ] { };
    \node[shape=circle,fill=gray] (J3) at (+0.1,+0.9) [nodeJ] { };
    \node[shape=circle,fill=gray] (J8) at (-0.6,+0.7) [nodeJ] { };
    \node[shape=circle,fill=gray] (J4) at (-0.8,+0.2) [nodeJ] { };
    \node[shape=circle,fill=gray] (J5) at (-0.7,-0.5) [nodeJ] { };
    \node[shape=circle,fill=gray] (J6) at (-0.0,-0.9) [nodeJ] { };
    \node[shape=circle,fill=gray] (J7) at (+0.6,-0.85) [nodeJ] { };
    \draw [very thick](I) -- (J1);
    \draw [very thick](I) -- (J2);
    \draw [very thick](I) -- (J3);
    \draw [very thick](I) -- (J8);
    \draw [very thick](I) -- (J4);
    \draw [very thick](I) -- (J5);
    \draw [very thick](I) -- (J6);
    \draw [very thick, dashed] (I) -- (J7);
    \draw [very thick](J1) -- (J2);
    \draw [very thick](J2) -- (J3);
    \draw [very thick](J3) -- (J8);
    \draw [very thick](J8) -- (J4);
    \draw [very thick](J4) -- (J5);
    \draw [very thick](J5) -- (J6);
    \draw [very thick](J6) -- (J7);
    \draw [very thick](J7) -- (J1);
    \end{tikzpicture}
  }
\hspace{0.5in}\\
\subfigure[A one-level stencil with only 4 direct neighbor vertices.]{
	\label{l1-tria4}
	\centering
	\begin{tikzpicture}[scale=1.5]
	\node[shape=circle,fill=gray] (I)  at (+0.0,+0.0) [nodeI] { };
	\node[shape=circle,fill=gray] (J1) at (+1.0,+0.0) [nodeJ] { };
	\node[shape=circle,fill=gray] (J2) at (+0.0,+1.0) [nodeJ] { };
	\node[shape=circle,fill=gray] (J3) at (-1.0,+0.0) [nodeJ] { };
	\node[shape=circle,fill=gray] (J4) at (-0.0,-1.0) [nodeJ] { };
	\node[shape=circle,fill=gray] (J5) at (+1.0,+1.0) [nodeK] { };
	\node[shape=circle,fill=gray] (J6) at (-1.0,+1.0) [nodeK] { };
	\node[shape=circle,fill=gray] (J7) at (-1.0,-1.0) [nodeK] { };
	\node[shape=circle,fill=gray] (J8) at (+1.0,-1.0) [nodeK] { };
	\draw [very thick](I) -- (J1);
	\draw [very thick](I) -- (J2);
	\draw [very thick](I) -- (J3);
	\draw [very thick](I) -- (J4);
	\draw [very thick](J1) -- (J2);
	\draw [very thick](J2) -- (J3);
	\draw [very thick](J3) -- (J4);
	\draw [very thick](J4) -- (J1);
	\draw [very thin, blue!20](J1) -- (J5);
	\draw [very thin, blue!20](J5) -- (J2);
	\draw [very thin, blue!20](J2) -- (J6);
	\draw [very thin, blue!20](J6) -- (J3);
	\draw [very thin, blue!20](J3) -- (J7);
	\draw [very thin, blue!20](J7) -- (J4);
	\draw [very thin, blue!20](J4) -- (J8);
	\draw [very thin, blue!20](J8) -- (J1);
	\draw [very thick, dashed] (I) -- (J5);
	\draw [very thick, dashed] (I) -- (J6);
	\draw [very thick, dashed] (I) -- (J7);
	\draw [very thick, dashed] (I) -- (J8);
	\end{tikzpicture}
}
  \quad
  \subfigure[A two-level stencil in a triangle mesh.]{
    \label{l2-tri}
    \centering
    \begin{tikzpicture}[scale=1.1]
      \node[shape=circle,fill=gray] (I)  at (+0.0,+0.0) [nodeI] { };
      \node[shape=circle,fill=gray] (J1) at (+0.8,-0.3) [nodeJ] { };
      \node[shape=circle,fill=gray] (J2) at (+0.7,+0.4) [nodeJ] { };
      \node[shape=circle,fill=gray] (J3) at (+0.1,+0.9) [nodeJ] { };
      \node[shape=circle,fill=gray] (J4) at (-0.8,+0.2) [nodeJ] { };
      \node[shape=circle,fill=gray] (J5) at (-0.7,-0.5) [nodeJ] { };
      \node[shape=circle,fill=gray] (J6) at (-0.0,-1.1) [nodeJ] { };
      \draw [very thick] (I) -- (J1);
      \draw [very thick] (I) -- (J2);
      \draw [very thick] (I) -- (J3);
      \draw [very thick] (I) -- (J4);
      \draw [very thick] (I) -- (J5);
      \draw [very thick] (I) -- (J6);
      \draw [very thick] (J1) -- (J2);
      \draw [very thick] (J2) -- (J3);
      \draw [very thick] (J3) -- (J4);
      \draw [very thick] (J4) -- (J5);
      \draw [very thick] (J5) -- (J6);
      \draw [very thick] (J6) -- (J1);
      %% J1
      \node[shape=circle,fill=gray] (J11) at (+1.4,+0.3) [nodeK] { };
      \node[shape=circle,fill=gray] (J12) at (+1.5,-0.5) [nodeK] { };
      \node[shape=circle,fill=gray] (J13) at (+1.1,-1.2) [nodeK] { };
      \draw [very thick] (J1) -- (J11);
      \draw [very thick] (J1) -- (J12);
      \draw [very thick] (J1) -- (J13);
      \draw [very thick] (J2) -- (J11);
      \draw [very thick] (J11) -- (J12);
      \draw [very thick] (J12) -- (J13);
      \draw [very thick] (J13) -- (J6);
      % J2
      \node[shape=circle,fill=gray] (J21) at (+1.0,+1.0) [nodeK] { };
      \draw [very thick] (J2) -- (J21);
      \draw [very thick] (J11) -- (J21);
      \draw [very thick] (J21) -- (J3);
      % J3
      \node[shape=circle,fill=gray] (J31) at (+0.4,+1.5) [nodeK] { };
      \node[shape=circle,fill=gray] (J32) at (-0.2,+1.6) [nodeK] { };
      \node[shape=circle,fill=gray] (J33) at (-0.9,+1.2) [nodeK] { };
      \draw [very thick] (J3) -- (J31);
      \draw [very thick] (J3) -- (J32);
      \draw [very thick] (J3) -- (J33);
      \draw [very thick] (J21) -- (J31);
      \draw [very thick] (J31) -- (J32);
      \draw [very thick] (J32) -- (J33);
      \draw [very thick] (J33) -- (J4);
      \draw [very thick, dashed] (J2) -- (J31);
      % J4
      \node[shape=circle,fill=gray] (J41) at (-1.6,+0.7) [nodeK] { };
      \node[shape=circle,fill=gray] (J42) at (-1.8,+0.0) [nodeK] { };
      \draw [very thick] (J4) -- (J41);
      \draw [very thick] (J4) -- (J42);
      \draw [very thick] (J33) -- (J41);
      \draw [very thick] (J41) -- (J42);
      % J5
      \node[shape=circle,fill=gray] (J51) at (-1.6,-1.2) [nodeK] { };
      \draw [very thick] (J5) -- (J42);
      \draw [very thick] (J5) -- (J51);
      \draw [very thick] (J42) -- (J51);
      %% J6
      \node[shape=circle,fill=gray] (J61) at (+0.2,-1.8) [nodeK] { };
      \node[shape=circle,fill=gray] (J62) at (-0.7,-1.7) [nodeK] { };
      \draw [very thick] (J6) -- (J61);
      \draw [very thick] (J6) -- (J62);
      \draw [very thick] (J61) -- (J62);
      \draw [very thick] (J13) -- (J61);
      \draw [very thick] (J62) -- (J5);
      \draw [very thick] (J51) -- (J62);
      \end{tikzpicture}
    }
  \caption{The stencils for divergence computing at a vertex (the red circle). The blue circles are the first level neighbor vertices, while the green circles are the second level neighbor vertices. }
  \label{}
\end{figure*}

%section 
\subsubsection{Construction of Divergence Stencils}
%%%% 以下讨论二维网格中的散度模板的构造

As described above, $f$ and $g$ are coupled in single linear system.
In order to obtain the two polynomials of $p$ degree, $p(p+3)$ conditions are required for 2D problems.
Each vertex gives two conditions, while each edge provides one condition.
In an unstructured mesh, for a vertex, it is easy to obtain all the direct neighbor vertices around it to construct a one-level stencil (Fig.~\ref{l1-tri} and \ref{l1-hyb}) which contains all the direct neighbor vertices and the edges (the thick lines in the figures) between them.
Sometimes, artificial edges can be created and included in the stencil if there are not enough edges in a one-level stencil. In Fig.~\ref{l1-hyb} and \ref{l1-tria4}, the dashed lines are artificial edges.

A one-level stencil in a triangular mesh contains $N$ neighbor vertices and $2N$ edges where $N=5$ or $6$, providing $4N$ conditions.
Thus, the accuracy order of a one-level stencil is between the 3rd and the 4th order.
To construct a stencil with higher order accuracy, we can simply combine the all one-level stencils of those direct neighbor vertices to construct a two-level stencil as in Fig.~\ref{l2-tri}.
This operation can be easily implemented with \verb|std::set| in C++ programming language.
It contains $18$ vertices and $42$ edges, providing $78$ conditions which should be enough for 2D polynomials higher than $6$ degree.

\subsubsection{Accuracy Tests of Divergence Computing}

\begin{figure}[hbtp]
  \centering
  \begin{minipage}{0.3\textwidth}
    \centering
    \includegraphics[width=\textwidth]{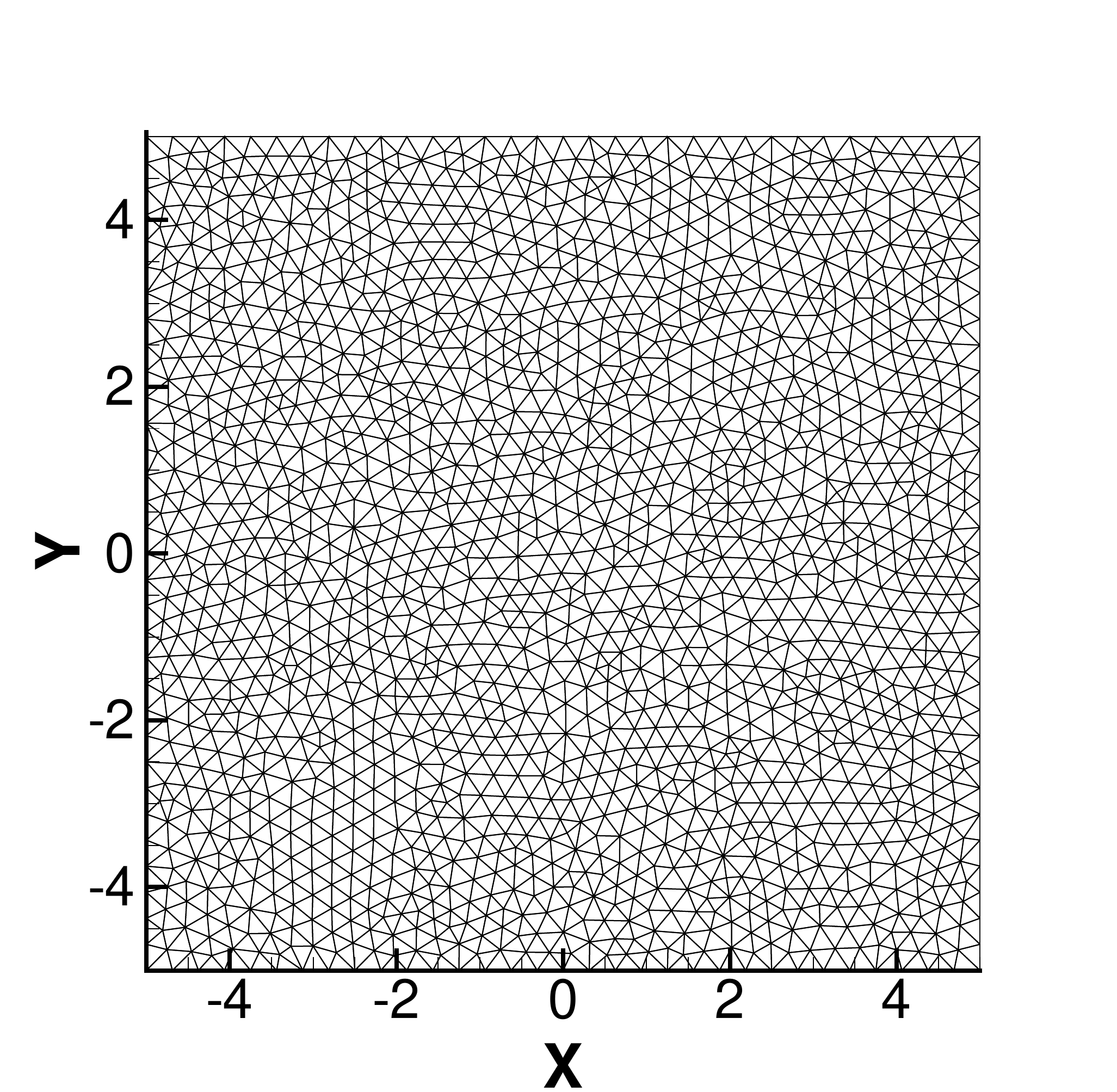}
    \caption{Unstructured mesh used to test divergence computing accuracy.}    
    \label{unsmesh}
  \end{minipage} \qquad
  \begin{minipage}{0.3\textwidth}
    \includegraphics[width=\textwidth]{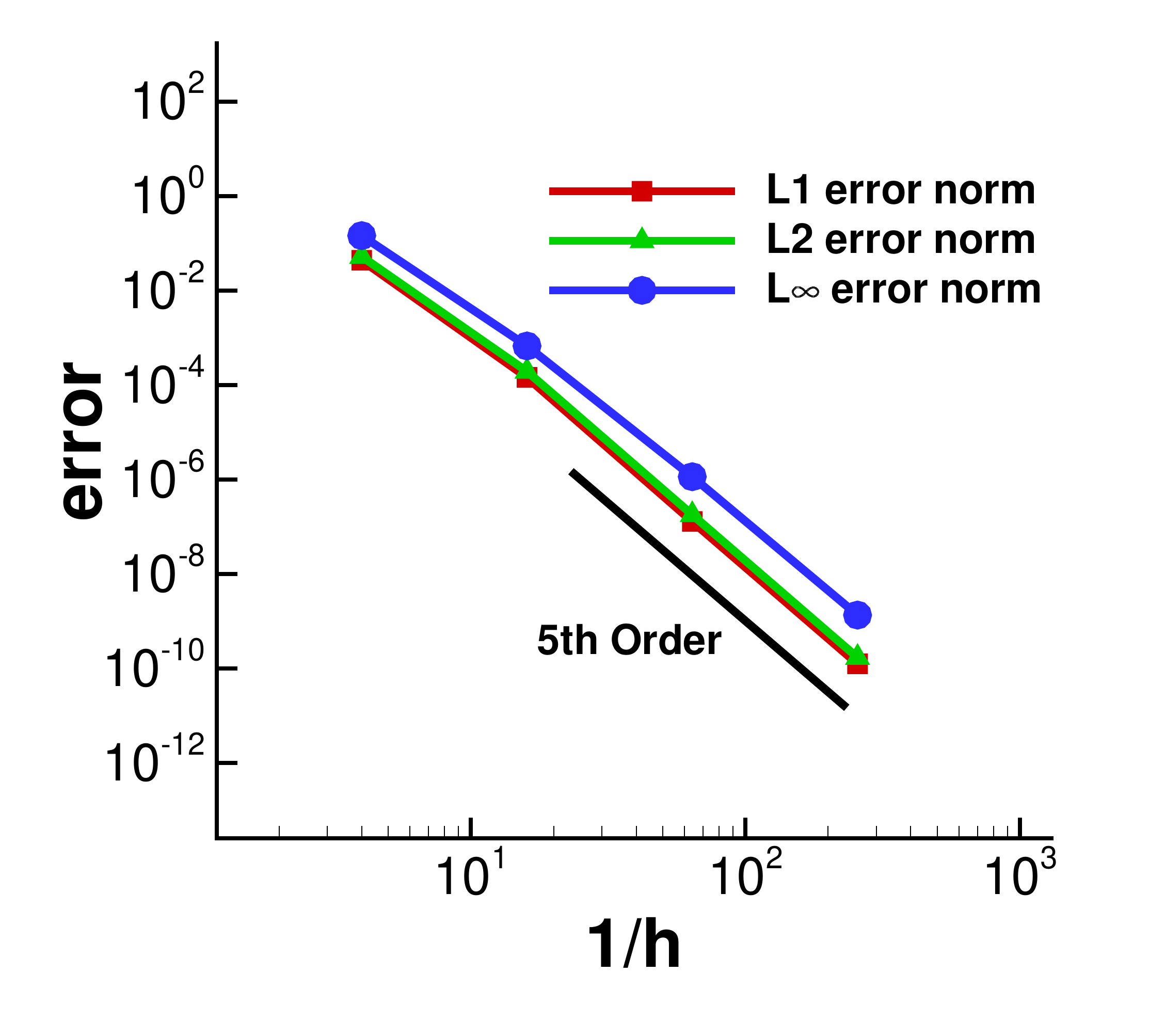}
    \caption{Three kinds of error norm of divergence computing accuracy test compared with the reference of fifth order.}
    \label{div_accuracy} 
    \centering
  \end{minipage}

\end{figure}

%%% 精度测试
In order to test the order of accuracy of the flux divergence computed by the least square method, a sequence of unstructured meshes similar to Fig.~\ref{unsmesh} with different mesh sizes are used. Two-level divergence stencils similar to Fig.~\ref{l2-tri} are utilized to support polynomial approximation of five degree, ie, fifth order accuracy of flux divergence computing. Exact flux vector and flux projection are assigned to mesh vertexes and mesh edges respectively, with which the flux divergence at vertexes are computed using the abovementioned least square method. The analytical solution for the test is given by
\begin{equation}
  \begin{cases}
    f(x) = sin(\frac{2\pi}{l}x)cos(\frac{2\pi}{l}y)\\
    g(x) = cos(\frac{2\pi}{l}x)sin(\frac{2\pi}{l}y).
  \end{cases}
\end{equation}
where $l$ is the length scale of the computation domain. Four boundaries are all set to periodic boundary condition. By comparing numerical results and the analytical divergence, three different kinds of error norms are evaluated. Along with the refinement of meshes, the decreasing tendency of all three error norms in Fig.~\ref{div_accuracy} show fifth order accuracy, which meets with our design expectation.
% A Monte Carol test was performed here.
% Many stencils are generated with random points around the origin.
% Some points serve as vertices, while others serve as middle points of edges with random directions.
% For each stencil, by scaling it to different sizes and approximating a known function, we can obtain the profile of $h \text{ to }\lVert\epsilon\rVert$ where $h$ is the length scale of the stencil and $\lVert\epsilon\rVert$ is the norm of approximating error.
% We plot all the  $h-\lVert\epsilon\rVert$ profiles of different random stencils to check how the order of accuracy varies with parameters such as number of vertices, number of edges, degree of the polynomials.

% 对于给定的多项式次数
% 随机生成许多模板，进行精度测试
% 将所有精度数据画在一个图上
% 不同 nnodes 和 nedges 配置，用不同的线型或者颜色区分
% 如果有必要，还比较法方程发发和SVD方法的差别

%subsection
\subsection{Computing Numerical Flux with One-Dimensional Stencil}
\label{comput_with_1d_stencil}
To achieve high order accuracy, interpolation or reconstruction based on high degree of polynomials is necessary. When using FDM in structured grids, interpolation or reconstruction is applied dimension by dimension.
On the contrary, in unstructured meshes, most methods apply interpolating/reconstruction directly in multidimensional space, which is much more complicated and expensive.
In addition, the operation should have WENO-like features so that discontinuities can be captured without oscillation. This means a number of candidate sub-stencils are considered and selected, making the operation even more complicated and expensive, especially in three-dimensional cases.
This can explain why FDM is one of the most efficient shock-capturing method among all available high order ones.
Herein, the author tried to achieve one-dimensional WENO interpolation in unstructured meshes so that the computing cost are largely reduced compared to other methods.

%subsubsection
\subsubsection{Construction of One-Dimensional Stencil in Unstructured Meshes}
\begin{figure}[htbp]
		\centering
		%\usetikzlibrary{shapes.geometric}
		\tikzstyle{nodeI}=[circle,draw=black!50, fill=black!60,thick, inner sep=0pt,minimum size=2mm]
		\tikzstyle{nodeE}=[rectangle,draw=blue!90, fill=white!60,thick, inner sep=0pt,minimum size=3mm]
		\begin{tikzpicture}[scale=1.]
		\node[shape=circle,fill=gray] (I)  at (+0.0,+0.0) [nodeI] { };
		\node[shape=circle,fill=gray] (J1) at (+0.8,-0.3) [nodeI] { };
		\node[shape=circle,fill=gray] (J2) at (+0.7,+0.4) [nodeI] { };
		\node[shape=circle,fill=gray] (J3) at (+0.1,+0.9) [nodeI] { };
		\node[shape=circle,fill=gray] (J4) at (-0.8,+0.2) [nodeI] { };
		\node[shape=circle,fill=gray] (J5) at (-0.7,-0.5) [nodeI] { };
		\node[shape=circle,fill=gray] (J6) at (-0.0,-1.1) [nodeI] { };
		\node[shape=circle,fill=gray] (E6) at (-0.0,-0.55) [nodeE] { };
		\draw [very thick] (I) -- (J1);
		\draw [very thick] (I) -- (J2);
		\draw [very thick, green] (I) -- (J3);
		\draw [very thick] (I) -- (J4);
		\draw [very thick] (I) -- (J5);
		\draw [very thick, red] (I) -- (J6);
		\draw [very thick] (J1) -- (J2);
		\draw [very thick] (J2) -- (J3);
		\draw [very thick] (J3) -- (J4);
		\draw [very thick] (J4) -- (J5);
		\draw [very thick] (J5) -- (J6);
		\draw [very thick] (J6) -- (J1);
		%% J1
		\node[shape=circle,fill=gray] (J11) at (+1.4,+0.3) [nodeI] { };
		\node[shape=circle,fill=gray] (J12) at (+1.5,-0.5) [nodeI] { };
		\node[shape=circle,fill=gray] (J13) at (+1.1,-1.2) [nodeI] { };
		\draw [very thick] (J1) -- (J11);
		\draw [very thick] (J1) -- (J12);
		\draw [very thick] (J1) -- (J13);
		\draw [very thick] (J2) -- (J11);
		\draw [very thick] (J11) -- (J12);
		\draw [very thick] (J12) -- (J13);
		\draw [very thick] (J13) -- (J6);
		% J2
		\node[shape=circle,fill=gray] (J21) at (+1.0,+1.0) [nodeI] { };
		\draw [very thick] (J2) -- (J21);
		\draw [very thick] (J11) -- (J21);
		\draw [very thick] (J21) -- (J3);
		% J3
		\node[shape=circle,fill=gray] (J31) at (+0.4,+1.5) [nodeI] { };
		\node[shape=circle,fill=gray] (J32) at (-0.2,+1.6) [nodeI] { };
		\node[shape=circle,fill=gray] (J33) at (-0.9,+1.2) [nodeI] { };
		\draw [very thick,green] (J3) -- (J31);
		\draw [very thick] (J3) -- (J32);
		\draw [very thick] (J3) -- (J33);
		\draw [very thick] (J21) -- (J31);
		\draw [very thick] (J31) -- (J32);
		\draw [very thick] (J32) -- (J33);
		\draw [very thick] (J33) -- (J4);
%		\draw [very thick, dashed] (J2) -- (J31);
		% J4
		\node[shape=circle,fill=gray] (J41) at (-1.6,+0.7) [nodeI] { };
		\node[shape=circle,fill=gray] (J42) at (-1.8,+0.0) [nodeI] { };
		\draw [very thick] (J4) -- (J41);
		\draw [very thick] (J4) -- (J42);
		\draw [very thick] (J33) -- (J41);
		\draw [very thick] (J41) -- (J42);
		% J5
		\node[shape=circle,fill=gray] (J51) at (-1.6,-1.2) [nodeI] { };
		\draw [very thick] (J5) -- (J42);
		\draw [very thick] (J5) -- (J51);
		\draw [very thick] (J42) -- (J51);
		%% J6
		\node[shape=circle,fill=gray] (J61) at (+0.2,-1.8) [nodeI] { };
		\node[shape=circle,fill=gray] (J62) at (-0.7,-1.7) [nodeI] { };
		\draw [very thick, green] (J6) -- (J61);
		\draw [very thick] (J6) -- (J62);
		\draw [very thick] (J61) -- (J62);
		\draw [very thick] (J13) -- (J61);
		\draw [very thick] (J62) -- (J5);
		\draw [very thick] (J51) -- (J62);
		%% J 62
		\node[shape=circle,fill=gray] (J621) at (+0.1,-2.6) [nodeI] { };
		\node[shape=circle,fill=gray] (J622) at (+1.0,-2.3) [nodeI] { };
		\draw [very thick,green] (J61) -- (J621);
		\draw [very thick] (J61) -- (J622);
		\draw [very thick] (J621) -- (J622);
		\draw [very thick] (J62) -- (J621);
		\draw [very thick] (J622) -- (J13);
		\end{tikzpicture}
	\caption{A one-dimensional interpolation stencil for the red edge in a triangular mesh. WENO5 interpolation is applied along this stencil to obtain flow states $\mathbf{w}_L$ and $\mathbf{w}_R$ at the middle point (the blue square) of the red edge.}
	\label{weno1d-tri}
\end{figure}
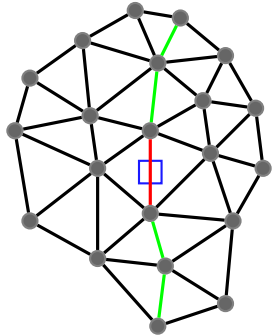

In unstructured meshes, there is no explicit ordered grid lines as in structural grids.
It is natural that people turn to multidimensional interpolation/reconstruction directly in each element (in DG, FR, and SD) or over several adjacent cells (in FVM).
However, we can still assemble a string as smooth as possible by assembling several connected edges. Fig.~\ref{weno1d-tri} shows such a string assembled by 5 edges. Along such a string, one-dimensional interpolation can be readily applied. Flow states $\mathbf{w}_L$ and $\mathbf{w}_R$ can be obtained by WENO5 interpolation at the middle point (the blue square) of the middle edge (the red edge).
After that, numerical flux at the middle point of the red edge can be computed with a Riemann solver.

When performing WENO interpolation along the one-dimensional stencil, there are two choices.
The first one is to interpolate in the generalized coordinate $\xi$.
In the $\xi$ domain, we assume that the length of the curve between two neighbor vertices is the same as $\Delta \xi=1$.
However, it is not guaranteed that the midpoint in $\xi$ domain is the same midpoint as on the original edge.
For nonuniform stencils constructed in unstructured meshes, the difference cannot be ignored.
%%%
In order to make the interpolation right at the middle point, another choice is interpolating along the curve length.
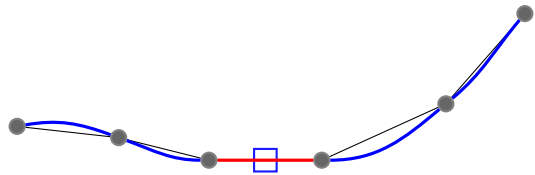
\begin{figure}
  \centering
  %\usetikzlibrary{shapes.geometric}
  \tikzstyle{nodeI}=[circle,draw=black!50, fill=black!60,thick, inner sep=0pt,minimum size=2mm]
  \tikzstyle{nodeE}=[rectangle,draw=blue!90, fill=white!60,thick, inner sep=0pt,minimum size=3mm]
  \begin{tikzpicture}[scale=1.5]
    \node  (L3) at (-2.2,+0.3) [nodeI] { };
    \node  (L2) at (-1.3,+0.2) [nodeI] { };
    \node  (L1) at (-0.5,+0.0) [nodeI] { };
    \node  (R1) at (+0.5,+0.0) [nodeI] { };
    \node  (R2) at (+1.6,+0.5) [nodeI] { };
    \node  (R3) at (+2.3,+1.3) [nodeI] { };
    \node  (E1) at (+0.0,+0.0) [nodeE] { };
    \draw [thin] (L3) -- (L2);
    \draw [very thick, blue] (L3) to[out=+10,in=-200] (L2);
    \draw [thin] (L2) -- (L1);
    \draw [very thick, blue] (L2) to[out=-20, in=-180] (L1);
    \draw [thin] (L1) -- (R1);
    \draw [very thick, red] (L1) -- (R1);
    \draw [thin] (R1) -- (R2);
    \draw [very thick, blue] (R1) to[out=0, in=-140] (R2);
    \draw [thin] (R2) -- (R3);
    \draw [very thick, blue] (R2) to[out=40, in=-130] (R3);
  \end{tikzpicture}
  \caption{A curve stencil(the color curve) genrated from linked edges(the thin black lines).}
  \label{curve_sten}
\end{figure}
As shown in Fig.~\ref{curve_sten}, two curves are constructed on each side of the middle edge.
From each end of the middle edge, a curve starts with the same tangent direction as the middle edge and cross the other two vertices.
Here, Catmull-Rom spline is adopted to construct the two curves.
The whole stencil is composed of the curves and a straight middle edge.
Along this stencil, nonuniform WENO5 interpolation can be applied along the curve length coordinate to obtain $\mathbf{w}_L$ and $\mathbf{w}_R$.

The one-dimensional interpolation used here is more efficient and can be applied on characteristics variables, making the solution more stable compared with interpolating conservative variables. Besides, the interpolation and flux computing methods used herein does not require a good-quality mesh as artificial edges can be used.

%subsubsection
\subsubsection{Accuracy Tests of One-Dimensional Stencil}
\label{sec_accuracy_test_1d_stencil}
Since interpolation stencils obtained from unstructured meshes are not equally spaced, non-uniform WENO scheme of fifth order accuracy is derived.
Details of the derivation is given in the appendix. The accuracy of the scheme is tested in this section. The same tests are conducted by using standard uniform WENO scheme for comparation. The exact solution is assumed to be a sinusoidal wave which reads
\begin{equation}
  f(x)=sin(x).
  \label{eq_sin}
\end{equation}

\begin{figure}[hbtp]
  \centering
  \subfigure[Uniform stencils]{
    \includegraphics[width=0.4\textwidth]{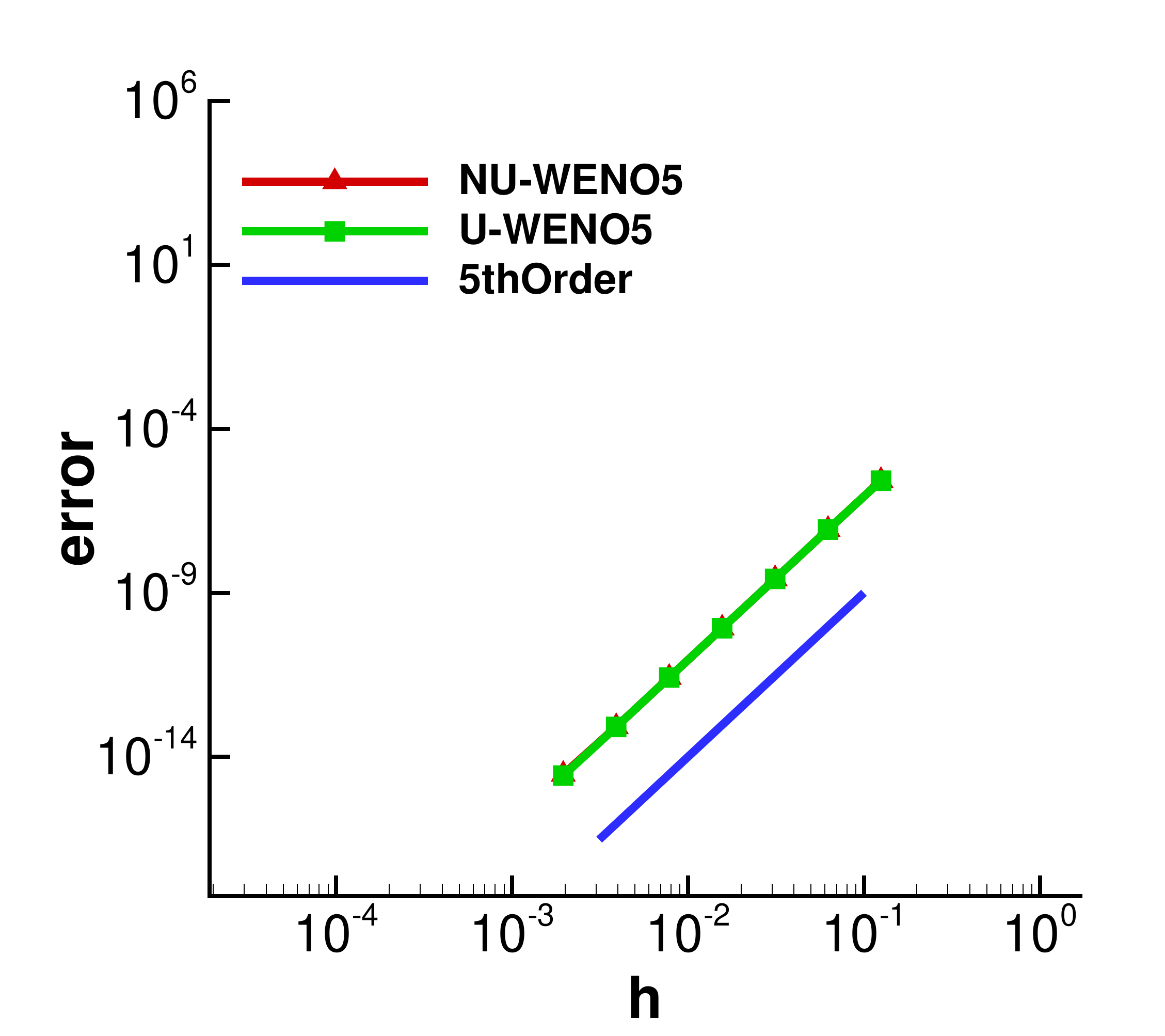}
  \label{nuweno_error_a}
  }\quad
  \subfigure[Non-uniform stencils]{
    \includegraphics[width=0.4\textwidth]{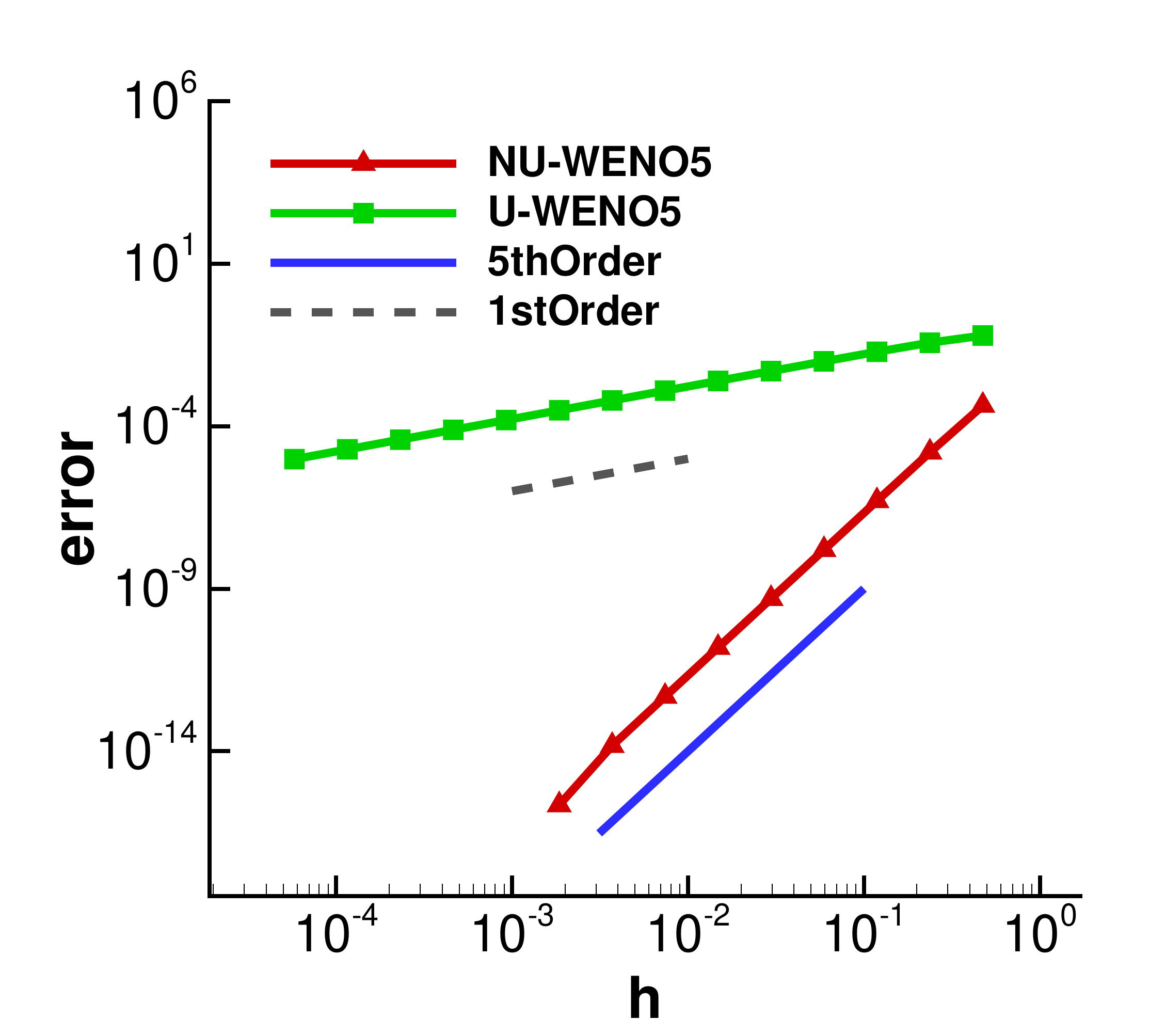}
  \label{nuweno_error_b}
  }
  \caption{L1 norm of the interpolation error of standard and non-uniform WENO schemes test with uniform and non-uniform stencils, where $\mathbf{h}$ is the interval between stencil nodes. NU-WENO5 stands for the non-uniform WENO scheme of fifth order of accuracy, while U-WENO5 means standard fifth order WENO scheme in reference \cite{jiang_efficient_1996}. }
  \label{nuweno_error}
\end{figure}

A stencil with five equally spaced points is the base stencil of the test. These points can be moved randomly to left or to right by some percent of the original interval. If the points do not move, standard and non-uniform WENO show the same order of accuracy as in Fig.~\ref{nuweno_error_a}, both presenting fifth order accuracy. By contrast, if points are moved randomly by 50 percent of original interval (which makes the stencil become non-uniform), non-uniform WENO scheme (NU-WENO5 in the figure) presents much better result compared with standard WENO scheme. Standard WENO5 scheme shows much larger interpolation error and has around first order of accuracy, while non-uniform WENO5 keeps almost the same small error as in the uniform-stencil case and preserves fifth order of accuracy (See Fig.~\ref{nuweno_error_b}). Therefore, it is necessary to use non-uniform WENO scheme if one intends to directly apply WENO schemes to non-uniform stencils without coordinate transformation.

\subsubsection{Accuracy Test Along Curves}
\label{sec_accuracy_test_2d_stencil}

Another two accuracy test cases are conducted to check the performance of standard WENO5 scheme and non-uniform WENO5 scheme along genuine curves. It is already shown in section \ref{sec_accuracy_test_1d_stencil} that non-uniform WENO5 scheme performs well in the case of one-dimensional stencil. In the following step, it is needed to test whether the scheme preserves the accuracy in the case of two-dimensional curves, which are extracted from unstructured meshes similar to the curve in Fig.~\ref{weno1d-tri}. Standard and non-uniform WENO5 schems will be used to interpolate the value at midpoint along these genuine curves.

\begin{figure}[hbtp]
  \centering
  \subfigure[Fifth-degree polynomial curve]{
    \includegraphics[width=0.4\textwidth]{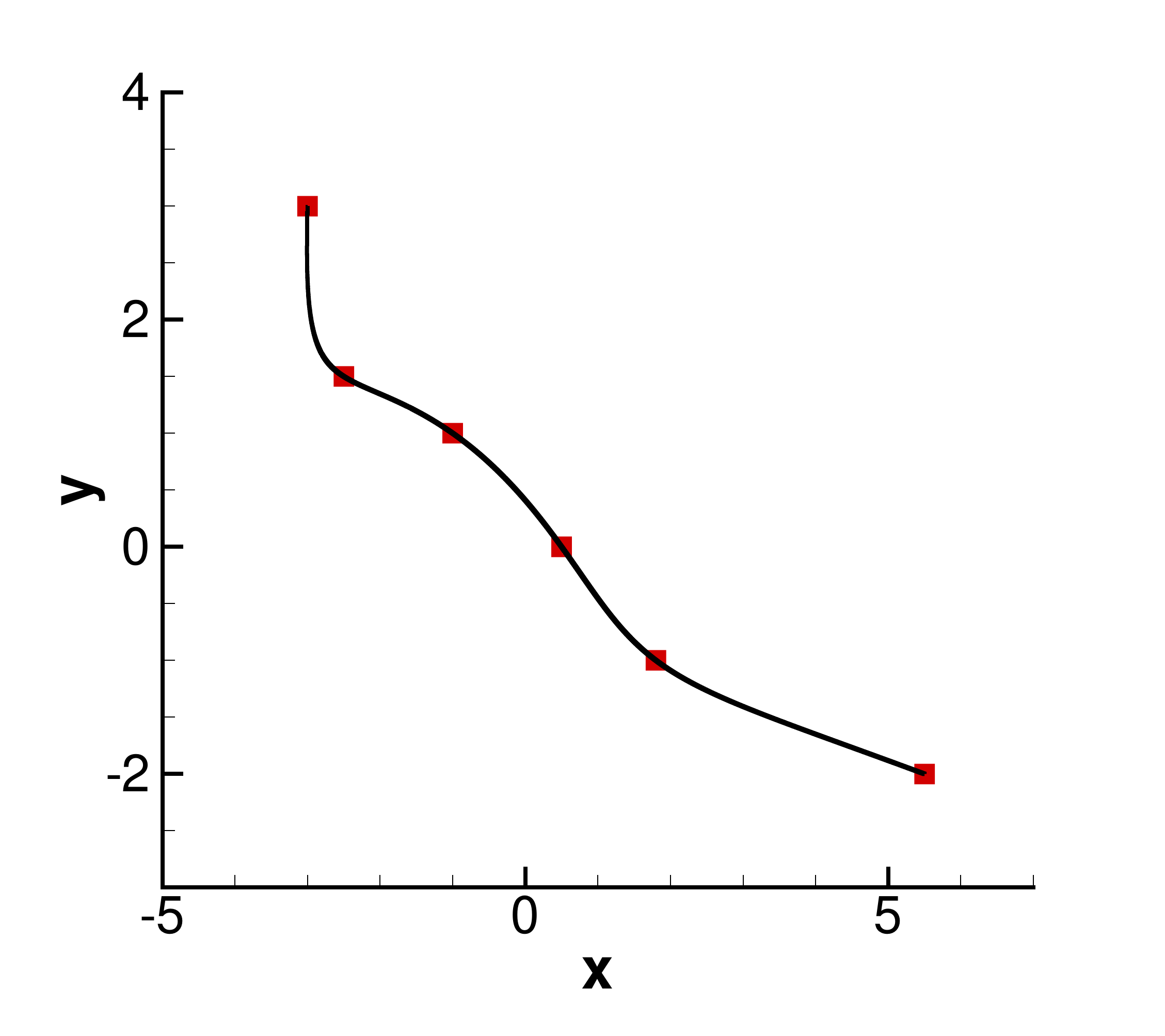}
    \label{nuweno_refine_curve}
  }\quad
  \subfigure[Interpolation error distribution]{
    \includegraphics[width=0.4\textwidth]{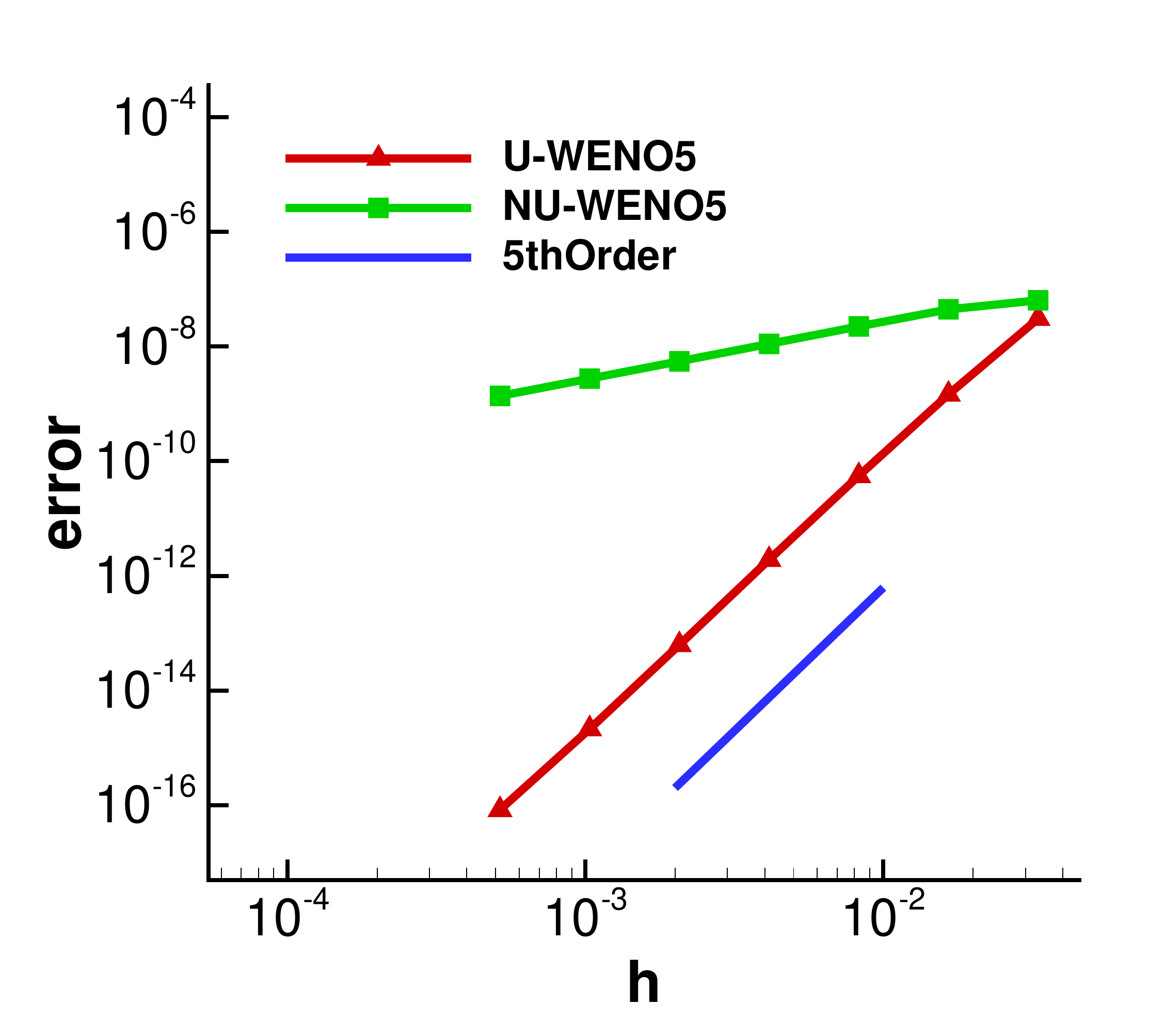}
    \label{nuweno_refine_error}
  }
  \caption{The interpolation curve and WENO interpolation results for the "refining" case. For the interpolation of the legend in this figure, the reader is referred to the figure legend in Fig.~\ref{nuweno_error}.}
  \label{nu_weno_curve_refine}
  
\end{figure}

\begin{figure}[hbtp]
  \centering
  \subfigure[Catmull-Rom curves]{
    \includegraphics[width=0.4\textwidth]{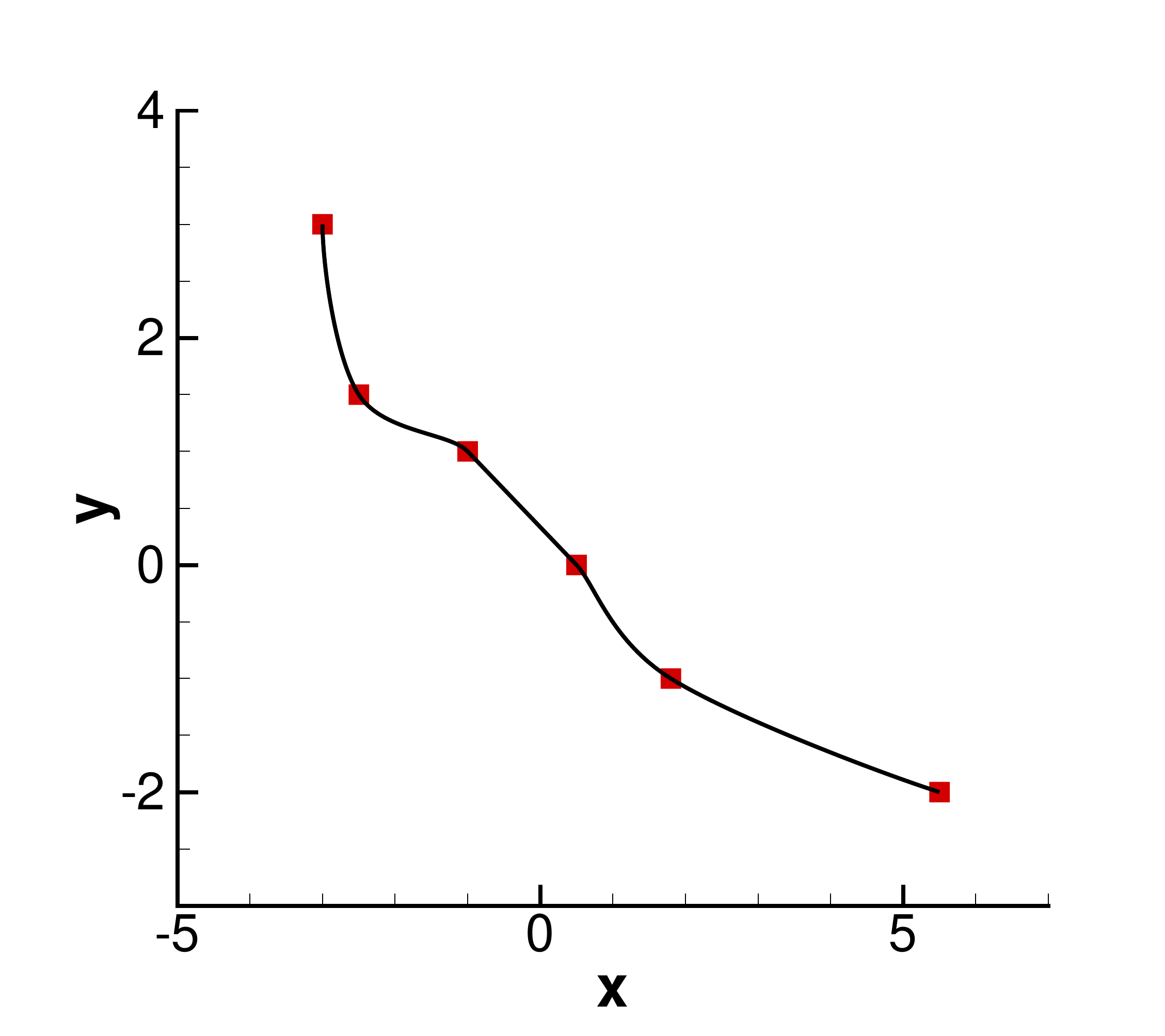}
    \label{nuweno_scale_curve}
  }\quad
  \subfigure[Interpolation error distribution]{
    \includegraphics[width=0.4\textwidth]{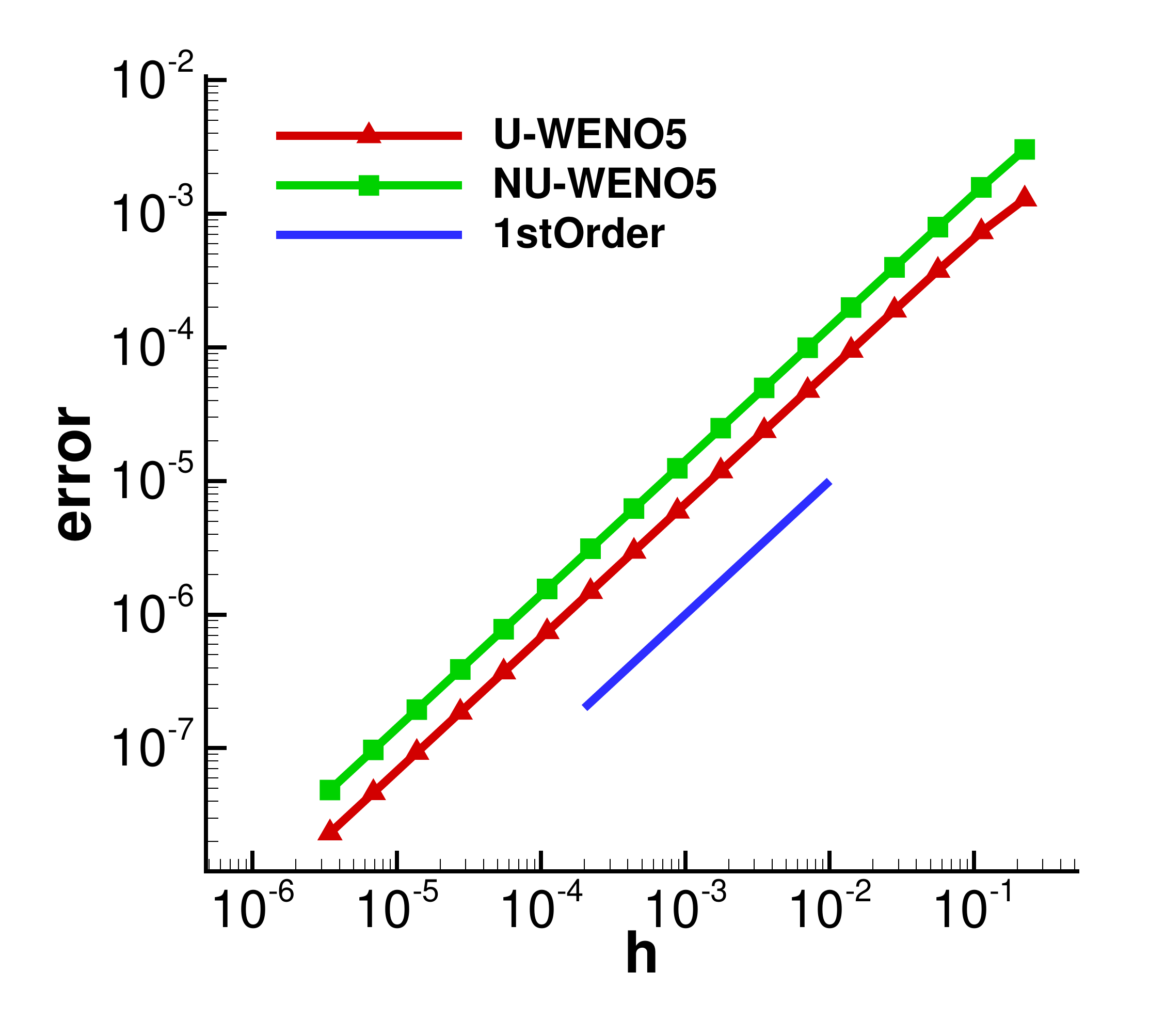}
    \label{nuweno_scale_error}
  }
  \caption{The interpolation curve and WENO interpolation results for the "scaling" case.}
  \label{nu_weno_curve_scale}
  
\end{figure}

Each curve in Fig.~\ref{nuweno_refine_curve} and Fig.~\ref{nuweno_scale_curve} is essentially a part of a two-dimensional field. Therefore, the test function used herein is two-dimensional rather than one-dimensional as in section \ref{sec_accuracy_test_1d_stencil}, which reads
\begin{equation}
  f(x,y)=sin(x)*cos(y)
  \label{eq_sin_2d}
\end{equation}

Fig.~\ref{nu_weno_curve_refine} shows the curve used in the first test case and the result of the accuracy test.
The curve in Fig.~\ref{nuweno_refine_curve} is obtained by fifth-degree polynomial interpolation that connects six nodes with $x=[{-}3.0\ {-}2.5\ {-}1.0\ 0.5\ 1.8\ 5.5],\ y=[3.0\ 1.5\ 1.0\ 0.0\ {-}1.0\ {-}2.0]$. Standard and non-uniform WENO5 schemes are utilized to do midpoint interpolation by using the known function values (values calculated by equation \ref{eq_sin_2d}) at nodes. This interpolation result is compared with the exact funciton value at the same midpoint to get the interpolation error. Then more nodes are added to the curve to do interpolation, thus making the average interpolation error much smaller. By refining the node distribution on the curve, we get a series of interpolation error as in Fig.~\ref{nuweno_refine_error}. It is found that non-uniform WENO5 scheme has the accuracy of fifth order when refining the node distribution, while the standard WENO5 scheme presents the accuracy of around first order. This result proves that non-uniform WENO5 scheme perserves fifth-order accuracy even along curves in two-dimensional filed if in a "refining" way. The basic reason for this performance lies in the fact that Taylor expansion of the non-uniform WENO5 gives the error of fifth-order, $O({\Delta x}^5)$. Therefore, with the refining of the node interval $\Delta x$, the interpolation error decrease in a fifth-order way.

In the second test case, it will be shown that WENO schemes no longer preserves the high accuracy if the test case is conducted in a "scaling" way rather than in a "refining" way. 

The curve in Fig.~\ref{nuweno_scale_curve} is a two-dimensional curve constructed by the Catmull-Rom method exactly in the same way as in Fig.~\ref{curve_sten}. It is naturally a part of a two-dimensional field and is extracted from an unstructured mesh. Unlike in the first test case, a series of interpolation error is obtained by scaling the curve and do the WENO interpolation, rather than refining the nodes on a curve. In the first test case, the function distribution along the whole curve will not change during the process of refinement. By contrast, the function value along the curve will change when the curve is scaled as it is not at the same spatial position, which makes the interpolation error unpredictable. The result in Fig.~\ref{nuweno_scale_error} shows an accuracy of first order. 

The above two test cases in fact correspond to the numerical accuracy test for the traditional structured grids, and for the method developed in this paper in the context of unstructured meshes respectively. When one is doing the accuracy test for the structured grids, the grid line in a dimension basically remains unchanged when the node distribution is refined, which exactly corresponds to the abovementioned "refining" case. However, for unstructured meshes, curves extracted from meshes as in Fig.~\ref{curve_sten} are much more like scaled rather than refined when the original unstructured meshes are refined. As a result, the accuracy test of the finite difference method developed in this paper for general unstructured meshes may have a similar performance as in Fig.~\ref{nu_weno_curve_scale}, which shows only around first-order accuracy. In fact, this conjecture will be tested and verified in the following paragraph. In section \ref{sec_isentropic_vortex}, the accuracy of spatial discretization error is tested on unstructured meshes. It is found that the discretization accuracy is of around first order for general unstructured meshes. Further details will be described in the corresponding section.

\section{Numerical Examples} \label{sec3}
Results of several inviscid problems are presented in this section to test the performance of the developed method. These cases include the problem with smooth solution and problems with discontinuities such as shock waves. 

In all of these cases, two layers of vertices and edges are used as the divergence stencil. One-dimensional six-point stencil is used for the fifth order WENO interpolation to compute numerical fluxes at the midpoints of edges. Thus, the overall spatial discretization accuracy is of the order fifth. The third order total variation diminishing Runge-Kutta scheme \cite{shu_efficient_1988} (RK-TVD) is used for time integration.

\subsection{Isentropic vortex problem}
\label{sec_isentropic_vortex}
The first test problem is from Yee et al. \cite{yee_low-dissipative_1999}, where an isentropic vortex convects in an inviscid free stream. The computation domain is of the size [-5,-5]$\times$[5,5], in which an isentropic vortex is located at [0,0] at the initial time. The vortex is added to a mean flow with $u_{\infty}=1$ and $v_{\infty}=1$, the initial flow field is given by
\[
\begin{aligned}
\rho =& \left[ 1-\frac{(\gamma-1)\beta^2}{8\gamma\pi^2} e^{1-r^2} \right]^{\frac{1}{\gamma-1}},\quad r^2=\bar{x}^2+\bar{y}^2,\\
(u,v) =& (1,1) + \frac{\beta}{2\pi}e^{\frac{1-r^2}{2}}(-\bar{y},\bar{x})\\
p=&\rho^{\gamma}
\end{aligned}
\]
where $\beta$ is the vortex strength and the value of 5.0 is used. Here, $(\bar{x},\bar{y})=(x-x_c, y-y_c)$, where $(x_c,y_c)$ is the center of the vortex at the initial time. The entire flow field is required to be isentropic, thus for a perfect gas, $p/\rho^\gamma=1$. Periodic boundary conditions are used in both directions, so that the vortex convects along the diagonal of the computational domain and reaches the initial position after each period.

\begin{figure}[hbtp!]
  \centering
  \includegraphics{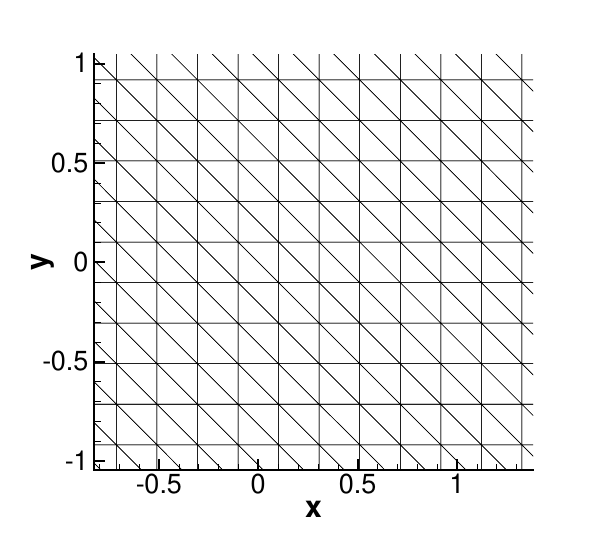}
  \caption{Uniform triangular mesh used in isentropic vortex case.}
  \label{isen_mesh}
\end{figure}

We use uniform triangular mesh which is obtained by diagonalizing the structured rectangular mesh of 50$\times$50, shown in Fig.~\ref{isen_mesh}.

\begin{figure}[hbtp!]
  \centering
  \subfigure[]{
    \label{isen_contour}
    \centering
    \includegraphics[width=0.4\textwidth]{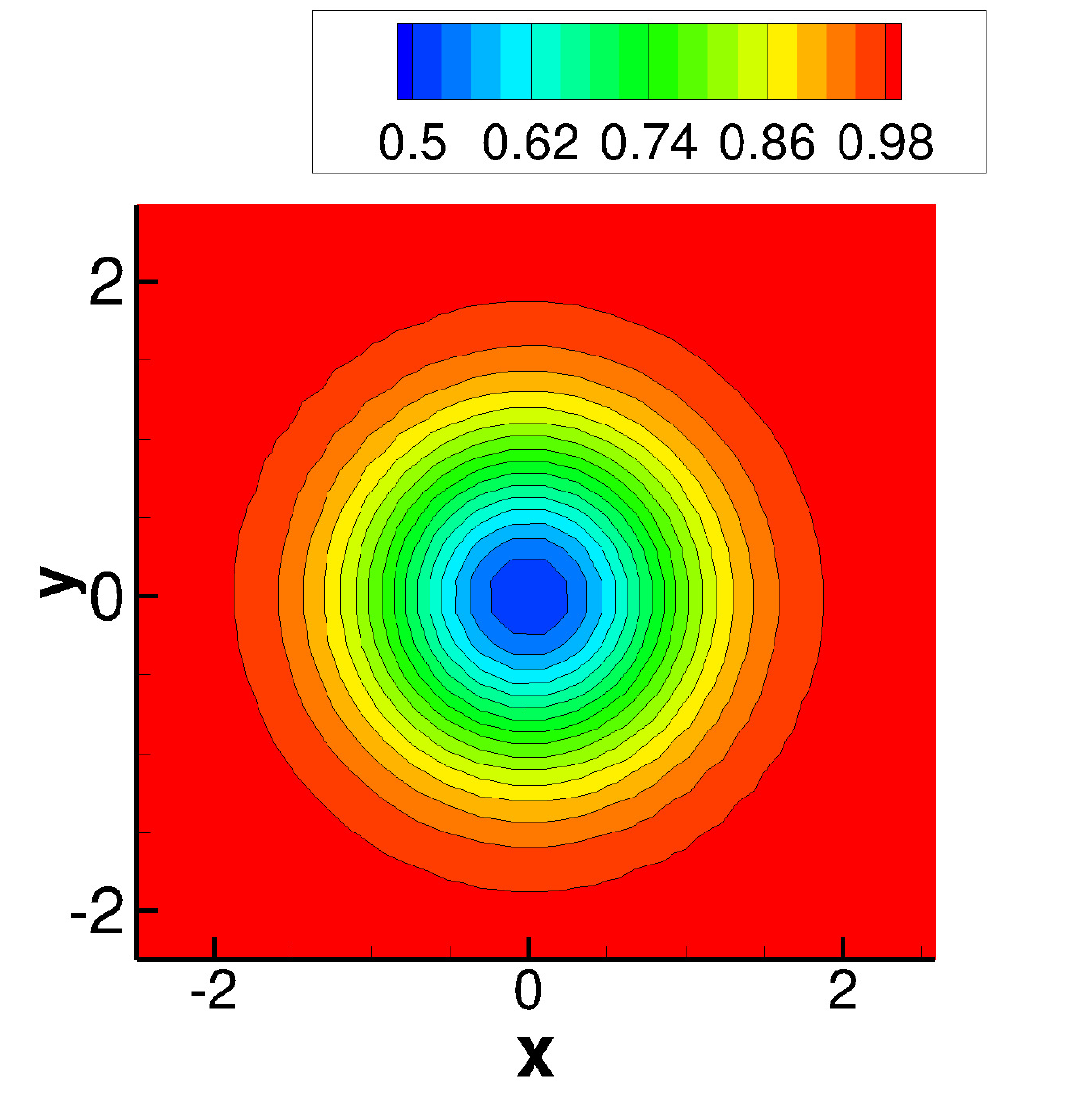}
  }
  \quad
  \subfigure[]{
    \label{isen_line}
    \centering
    \includegraphics[width=0.4\textwidth]{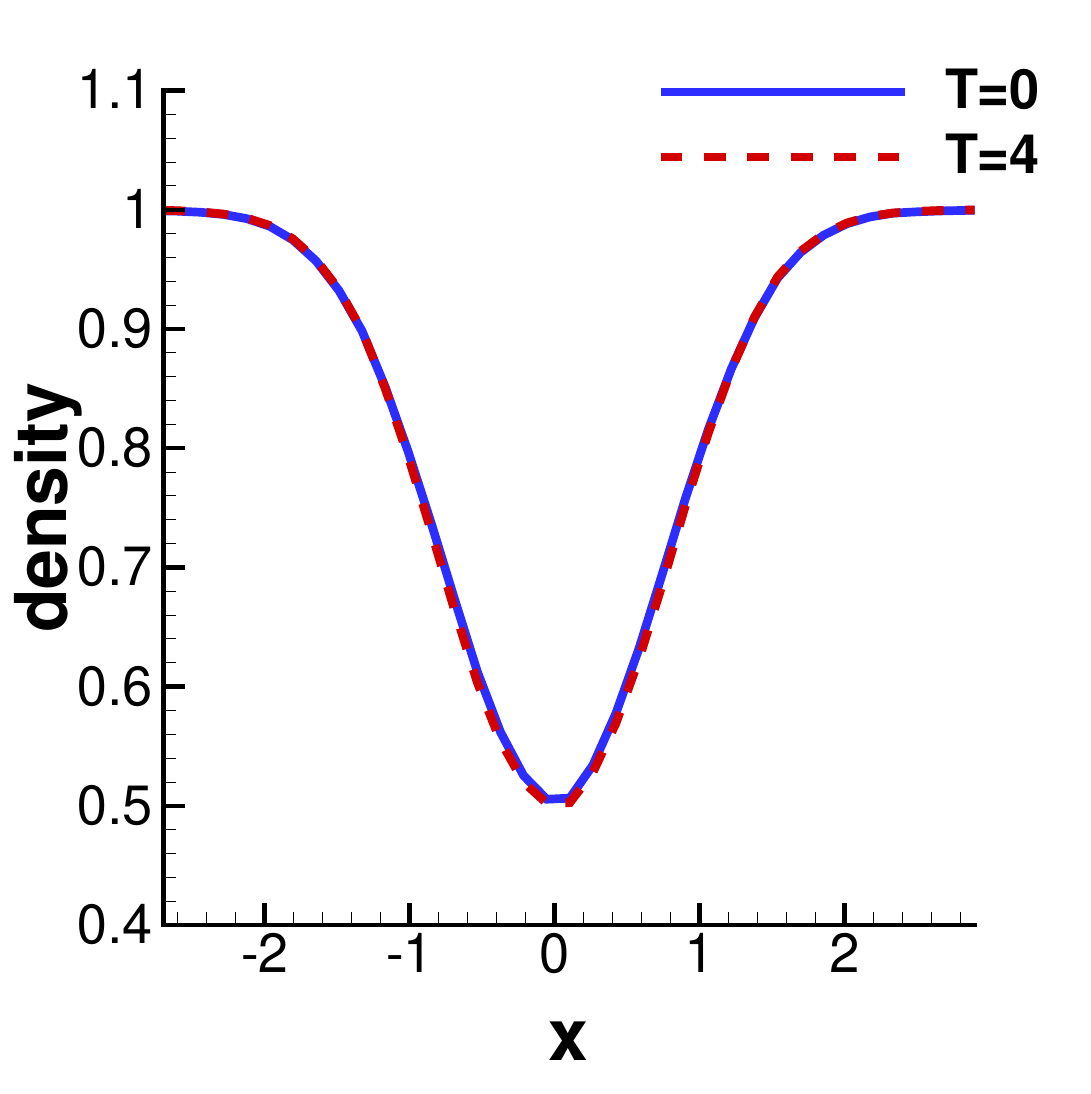}
  }
  \caption{Isentropic vortex advection at T=4 periods: (a) 17 density contours from 0.5 to 0.98; (b) density distribution at the line $y=0$.}
  \label{isen_res}
\end{figure}

Fig.~\ref{isen_res} gives the results of calculation after four time periods. Density contours are presented in Fig.~\ref{isen_contour} and it's distribution at the cut line $y=0$ is shown in Fig.~\ref{isen_line} at the same time. It can be seen that after four time periods, the vortex arrives at the initial position and remains almost the same distribution compared with the initial flow field, which indicates very low dissipation. 

\begin{figure}
  \centering
  \includegraphics[width=0.4\textwidth]{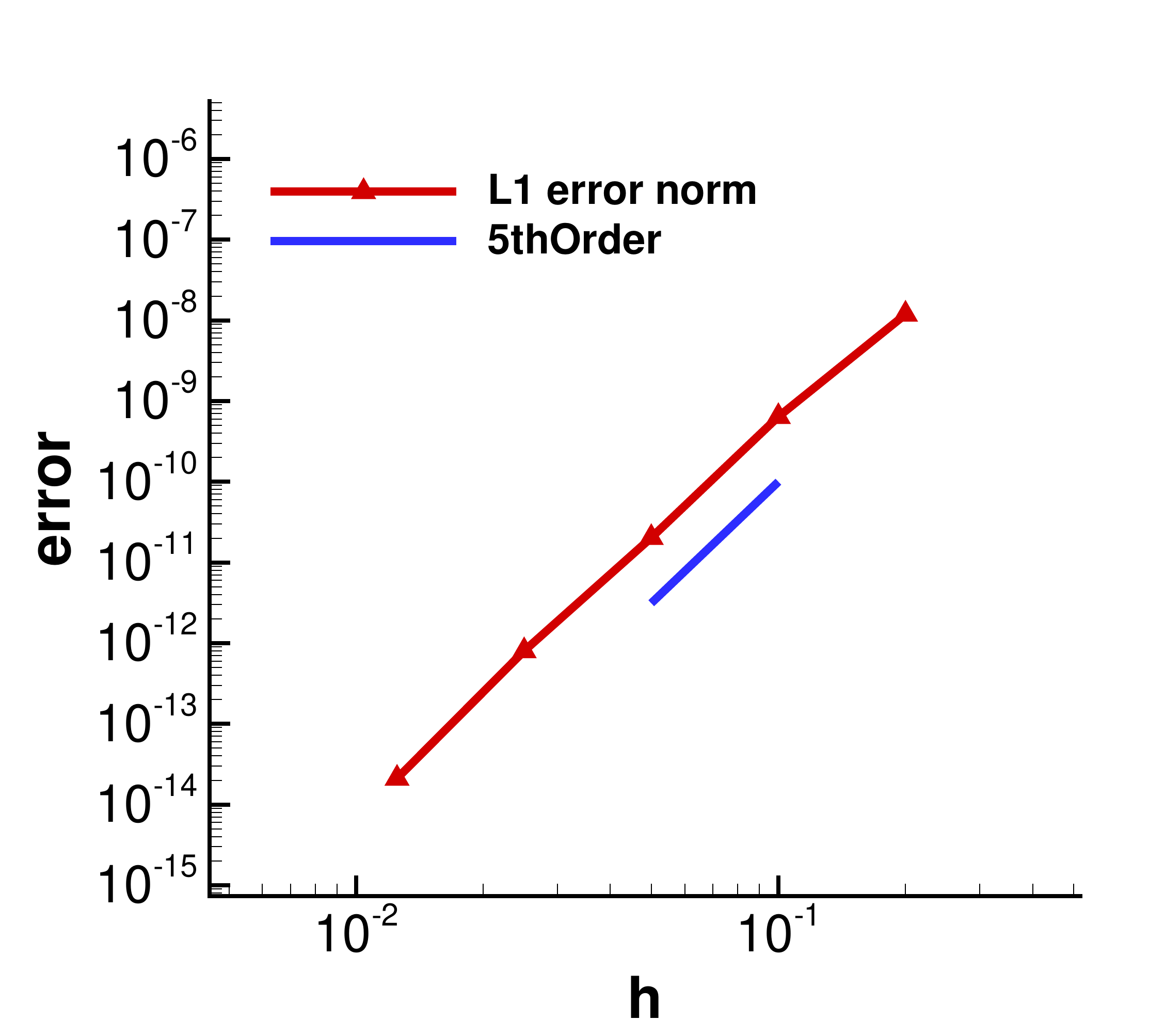}
  \caption{Accuracy test of the isentropic vortex problem on regular triangular meshes.}
  \label{isen_accuracy}
\end{figure}

A sequence of triangular meshes similar to Fig.~\ref{isen_mesh} with different resolutions are used to test the spatial discretization accuracy of the present method. To remove the error caused by time discretization, the computational error of the density is recorded after just one small time step where CFL (Courant-Friedrichs-Lewy condition) number is set to 0.01. The result in Fig.~\ref{isen_accuracy}, in terms of the $L_1$ error norm of density, shows that the present method has the accuracy of fifth order on regular triangular meshes as in Fig.~\ref{isen_mesh}, which agrees with the design expectation. 

\begin{figure}
  \centering
  \includegraphics[width=0.4\textwidth]{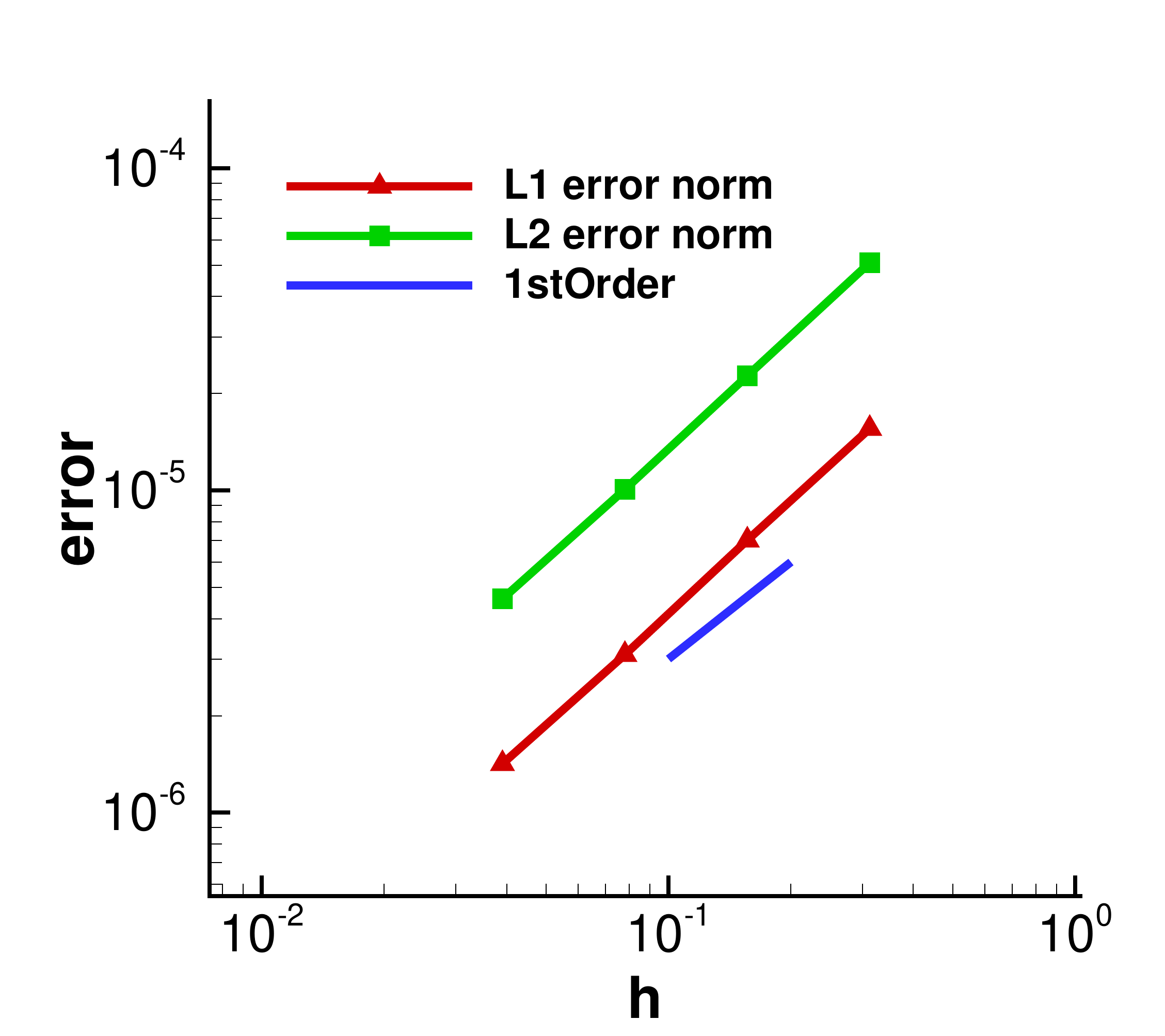}
  \caption{Accuracy test of the isentropic vortex problem on general triangular meshes.}
  \label{fig_isen_general}
\end{figure}

Another sequence of general triangular meshes generated by Delaunay method are used to test the discretization accuracy. The unstructured meshes are similar to the mesh in Fig.~\ref{unsmesh}, but are refined in order while keeping the computation domain unchanged as $x\in[-5,5],\ y\in[-5,5]$. By using the developed spatial discretization method, the isentropic vortex is convected only in only one small time step with CFL=0.01. The simulation results are compared to the exact solution and L1 norm and L2 norm of density error are recorded. The results in Fig.~\ref{fig_isen_general} show the accuracy of around first order. The reason for this low order accuracy lies in the way that unstructured meshes are refined during the test. When those general unstructured meshes are refined, the interpolation stencils/curves on them are scaled rather than refined, which leads to the same situation as in Fig.~\ref{nu_weno_curve_scale} in section \ref{sec_accuracy_test_2d_stencil} and is only first-order accurate. How to improve the performance of the accuracy in the case of general unstructured mesh is left to the future work. Herein the focus of the study lies in the idea of constructing one-dimensional interpolation stencil on unstructured mesh. It shows the potential to improve the accuracy and efficiency of the simulation with unstructured meshes and to be applied to cases with solution discontinuity.

%subsection
\subsection{Two-Dimensional Shock Tube}
\label{sec_2d_Sod}
\begin{figure}[hbtp!]
  \centering
  \includegraphics[width = 0.4\textwidth]{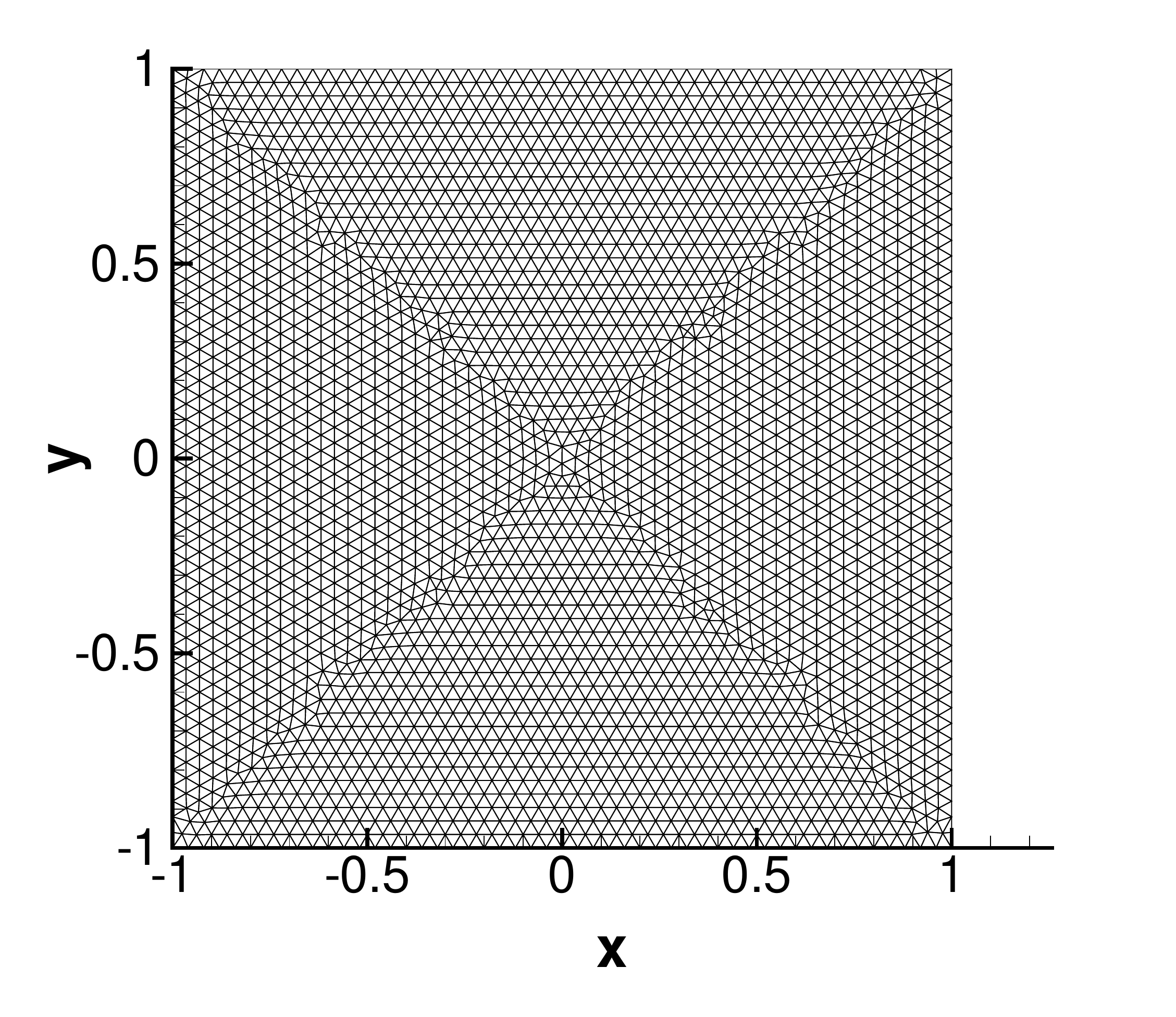}
  \caption{The triangular mesh used for 2d Sod shock tube case.}
  \label{sod_mesh}
\end{figure}

This problem is an extension of the one-dimensional shock tube problem introduced by Sod \cite{sod1978survey}, where high pressure and high density are set a small region at the initial time. The computational domain has size $x\in[-1,1]\, y\in[-1,1]$ and periodic boundary conditions are used in both directions. Front advancing method is used to generate the triangular mesh. The average mesh size is around 0.04 (with equally spaced 51 points on each boundary), and there are 3058 solution points in total, see Fig.~\ref{sod_mesh}.

\begin{figure}[hbtp]
  \centering
  \subfigure[t=0.0]{
    \includegraphics[width =0.4\textwidth]{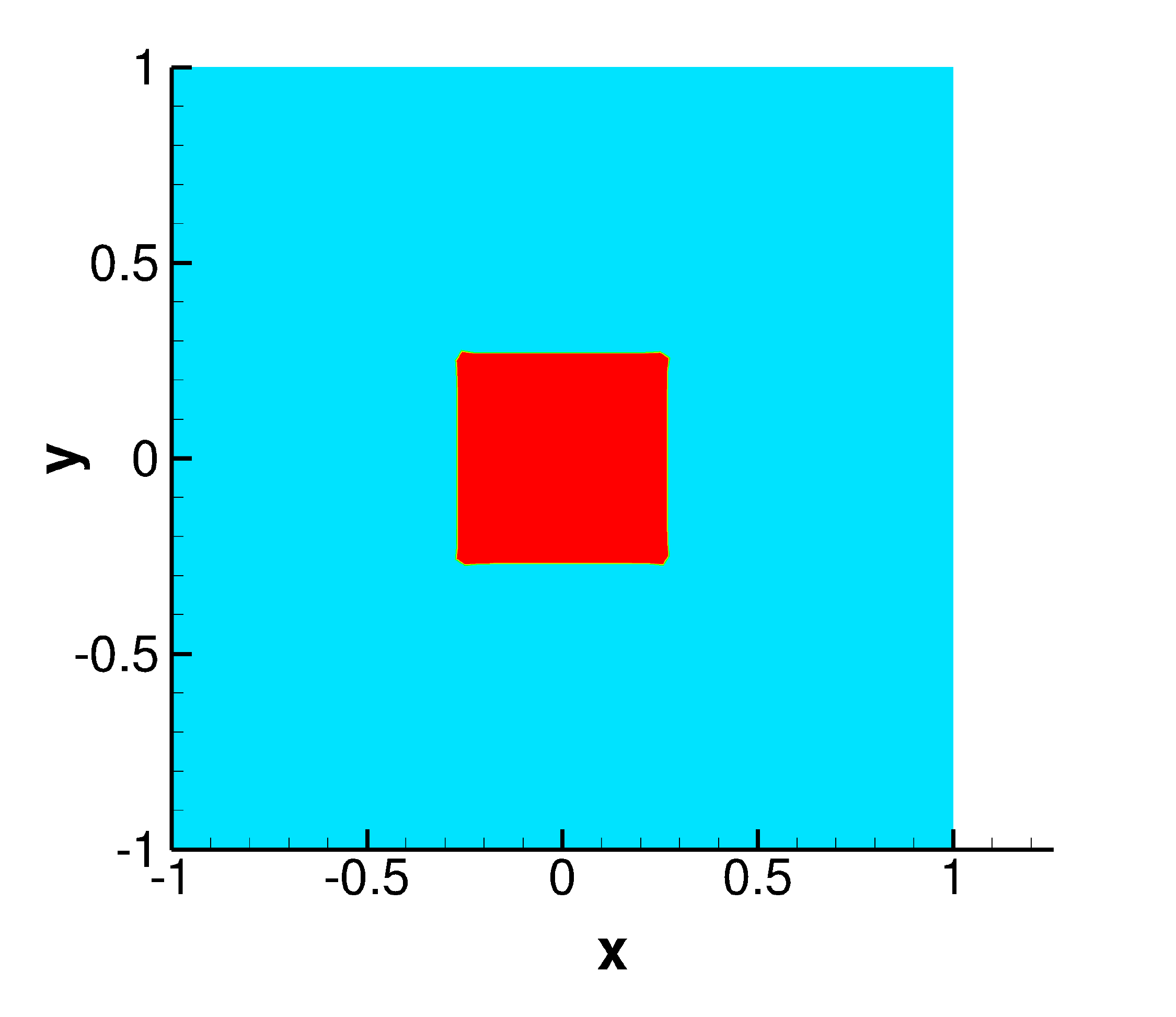}
    \label{2d_sod0}
  }
  \subfigure[t=0.2]{
    \includegraphics[width =0.4\textwidth]{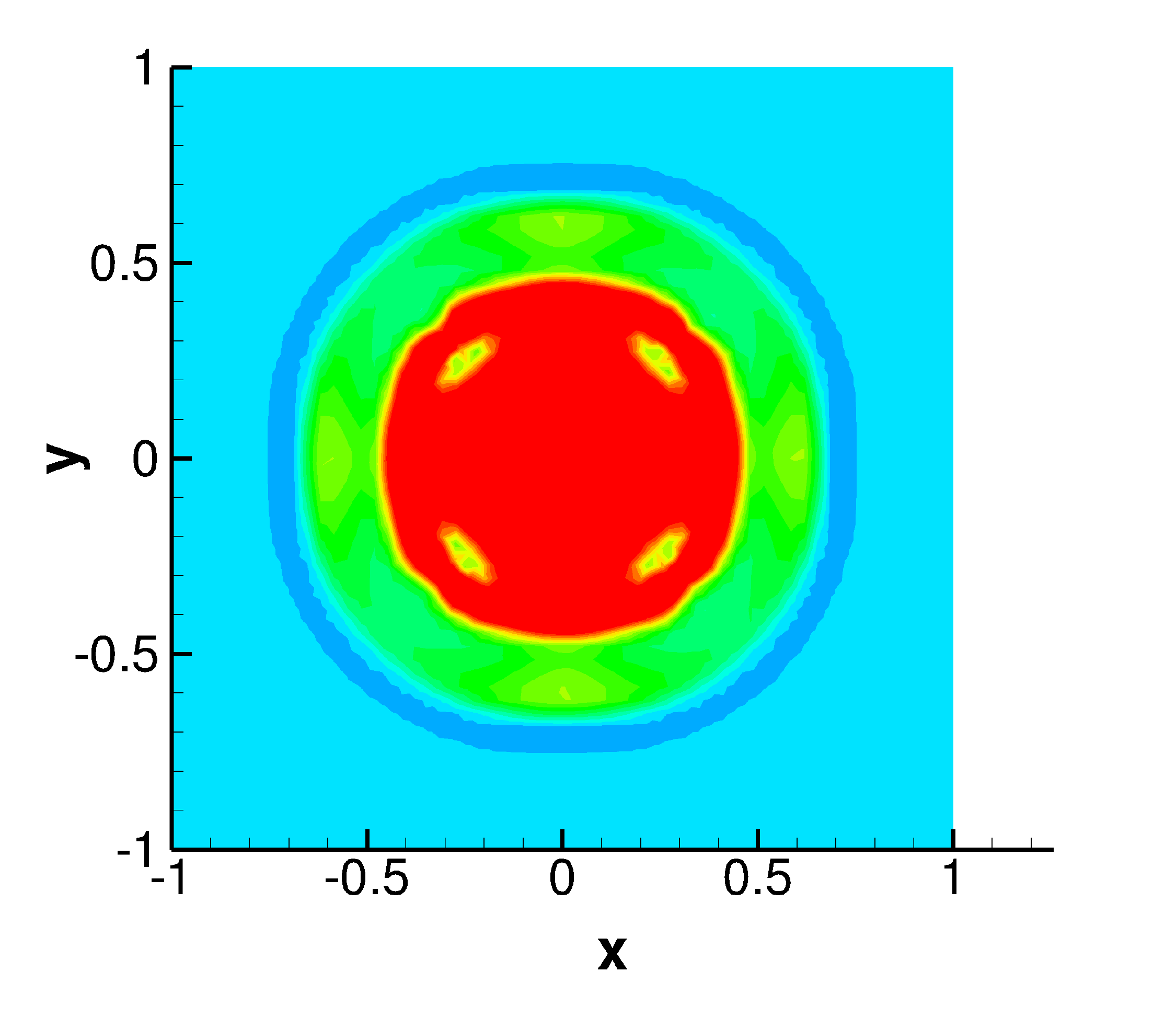}
    \label{2d_sod2}
  }
  \subfigure[t=0.4]{
    \includegraphics[width =0.4\textwidth]{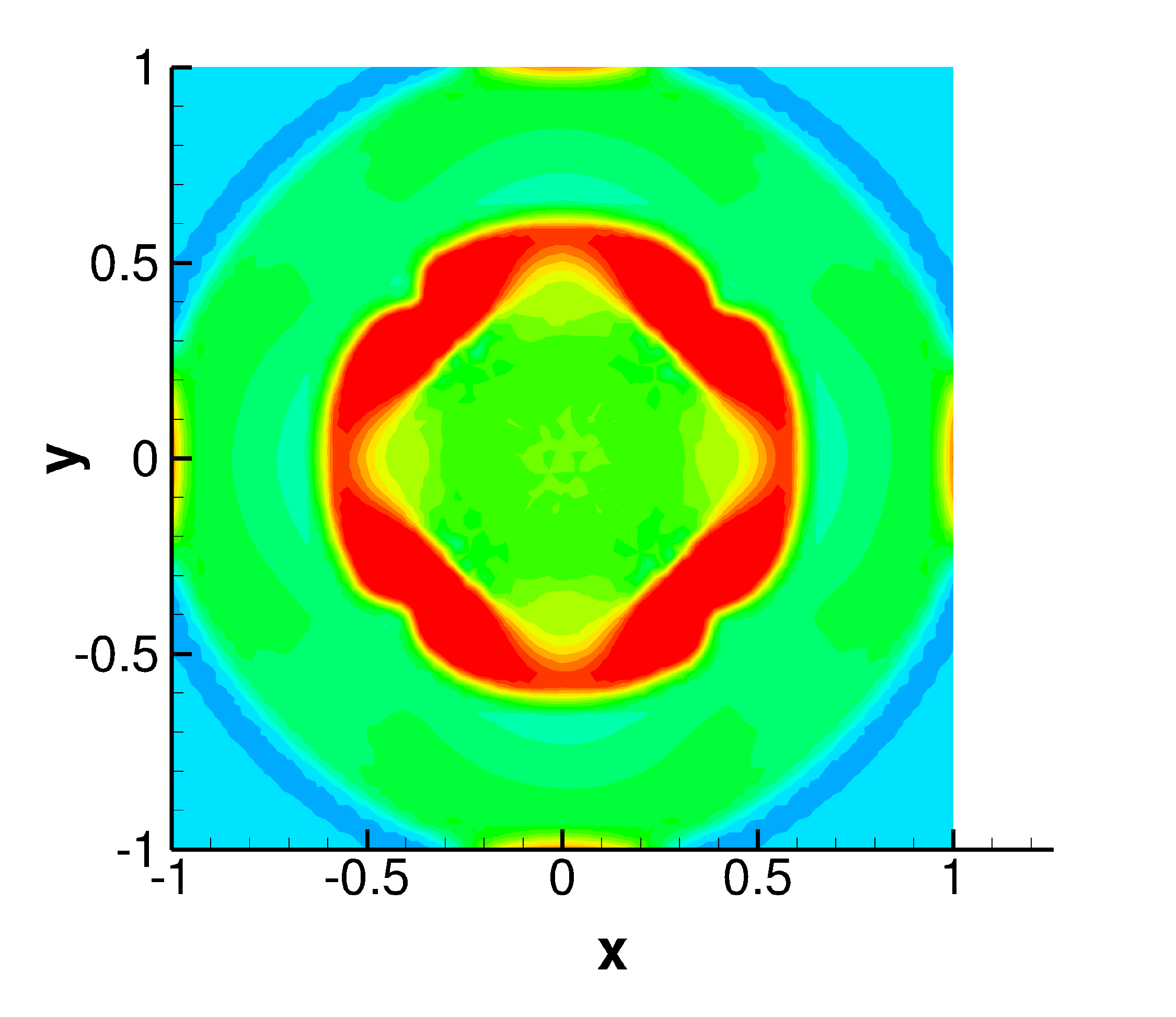}
    \label{2d_sod4}
  }
  \subfigure[t=0.6]{
    \includegraphics[width =0.4\textwidth]{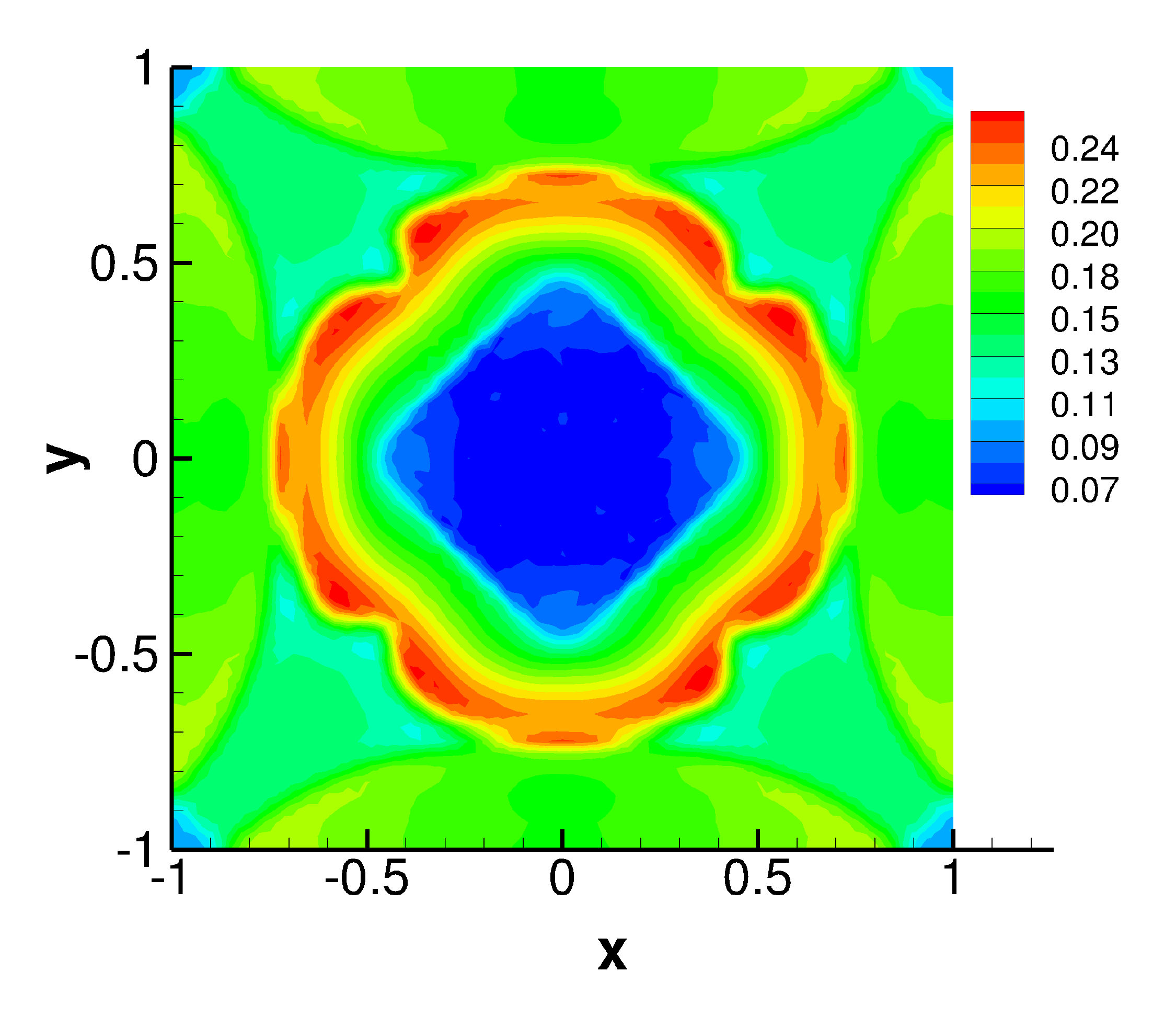}
    \label{2d_sod6}
  }
  \caption{Density contours from 0.07 to 0.25 of two-dimensional Sod shock tube problem from time $t=0$ to $t=0.6$.}
  \label{2d_sod}
\end{figure}

\begin{figure}
  \subfigure[]{
    \label{2dsod_comp_a}
    \includegraphics[width=0.3\textwidth]{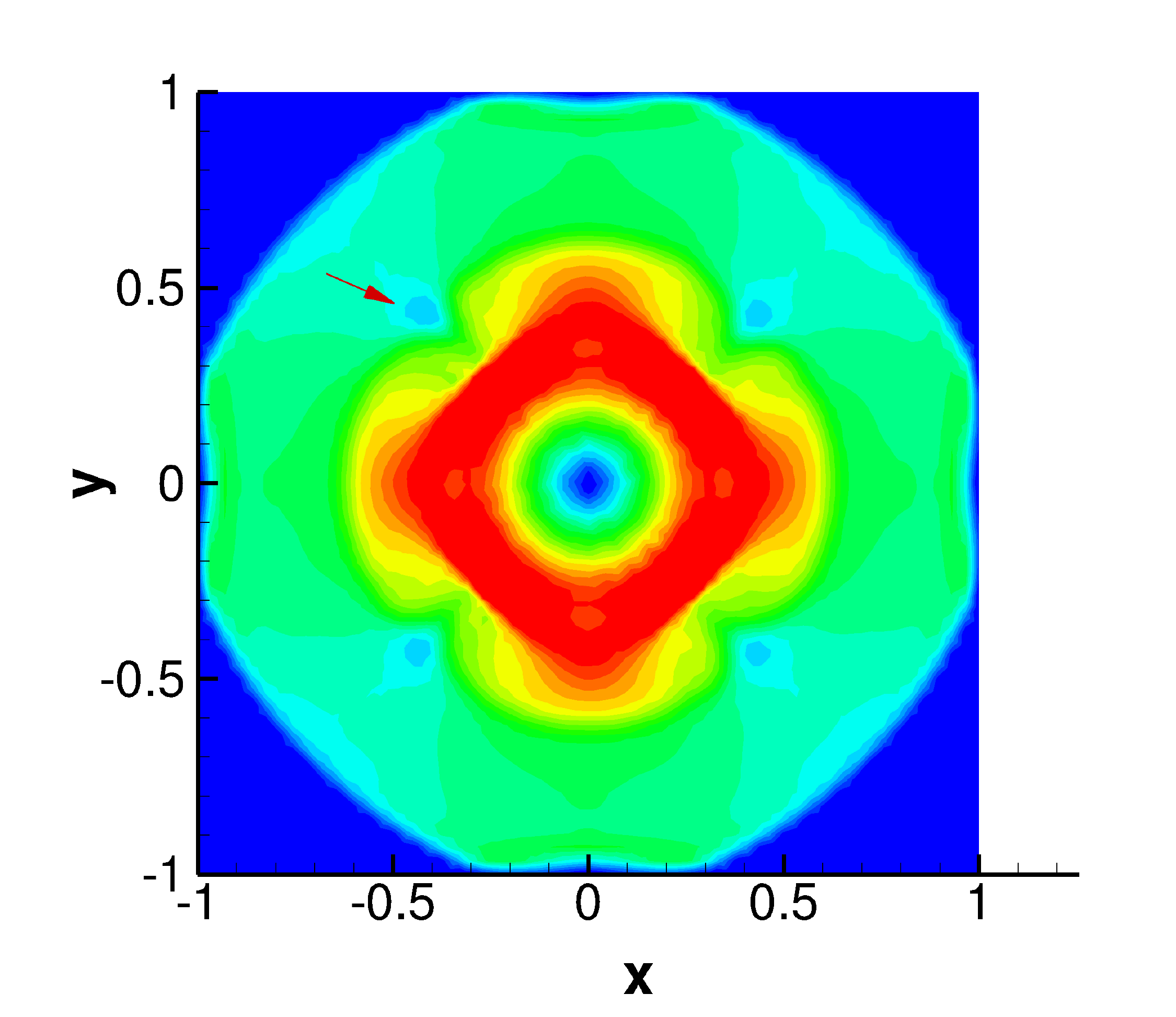}
  }
  \subfigure[]{
    \label{2dsod_comp_b}
    \includegraphics[width=0.285\textwidth]{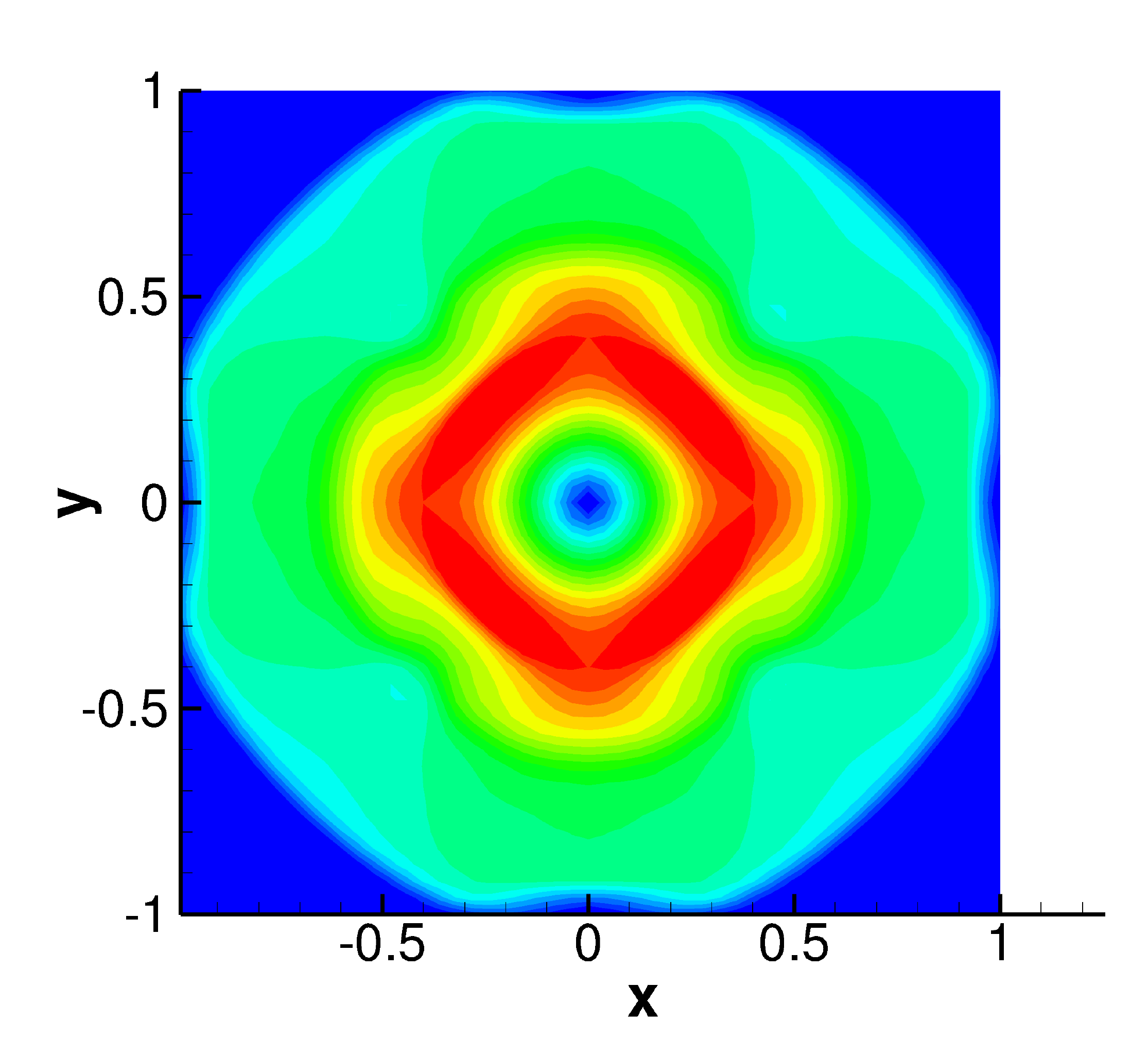}
  }
  \subfigure[]{
    \label{2dsod_comp_c}
    \includegraphics[width=0.3\textwidth]{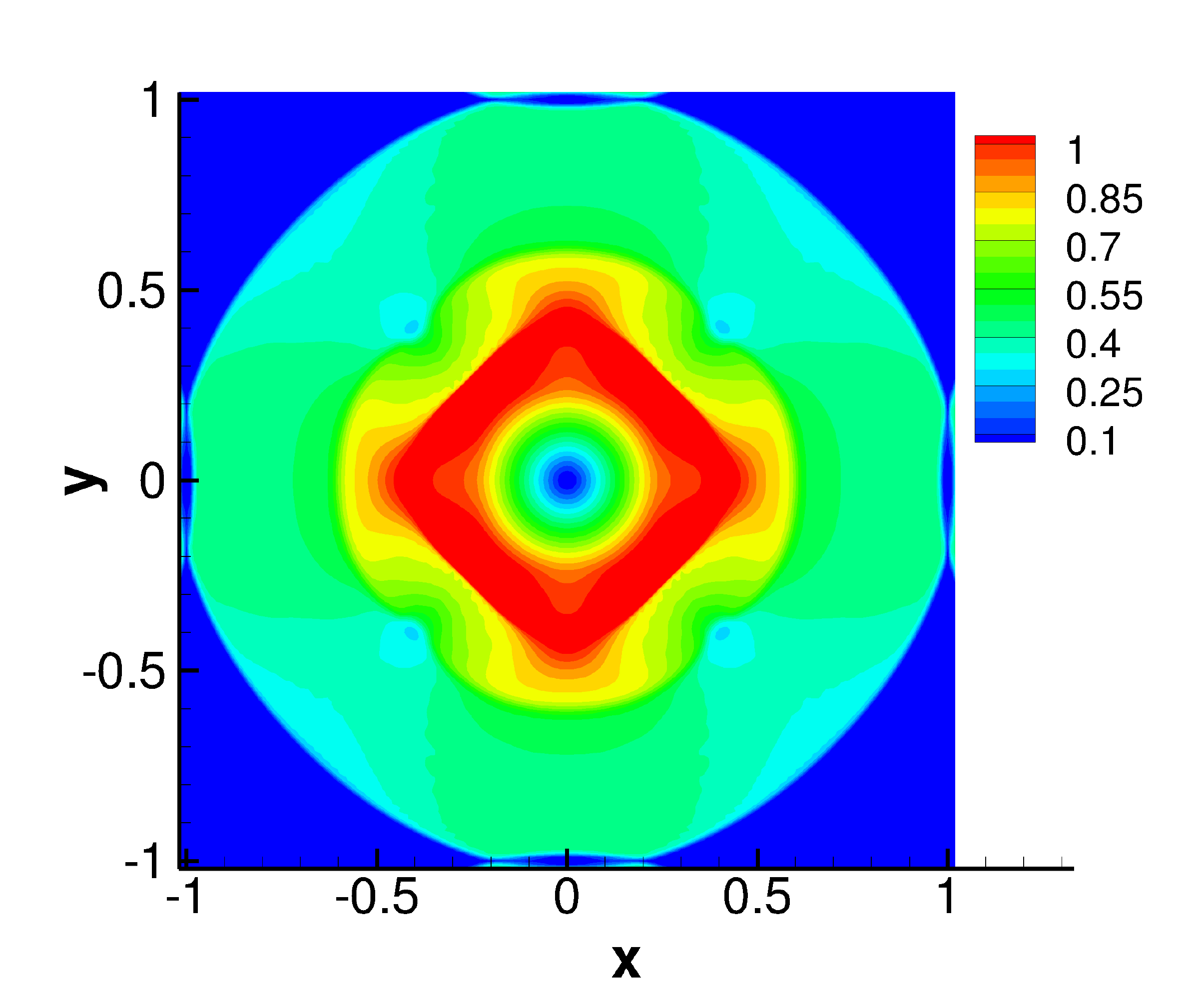}
  }
  \centering
  \caption{19 equally spaced Mach number contours from 0.1 to 1.0, at $t=0.4$: (a) unstructured mesh with 3058 solution points; (b) structured grid of the size $55\times55$; (c) structured grid of the size $201\times201$.}
  \label{2dsod_comp}
\end{figure}

The initial condition is given by
\[
  (\rho,u,v,p)=\begin{cases}
    (1,0,0,1),&|x|<0.25,|y|<0.25\\
    (0.125,0,0,0.1),&\mathrm{elsewhere}
  \end{cases}
\]
The ratio of specific heats $\gamma$ is set to 1.4. With the time evolution, shock waves, contact discontinuities and expansion fans are generated. Fig.~\ref{2d_sod} shows the evolution of density at four different moments, specifically, from $t=0$ to $t=0.6$ with an interval of $dt=0.2$. The solutions clearly capture shock waves and contact discontinuities, but with a little overshoot, which may be caused by insufficient dissipation.
To validate the solution, our results are compared with ones computed by WCNS scheme on structured grids, where midpoint-and-node-to-node difference (MND) scheme with sixth order accurate is used, which reads
\begin{equation}
  \begin{aligned}
  \left. \frac{\partial F}{\partial x} \right|_{i,j}= &\frac{1}{\Delta x}
  \left[
  \frac{3}{2}\left(\tilde{F}_{i+1/2,j}-\tilde{F}_{i-1/2,j}\right) - \frac{3}{10}\left(F_{i+1,j} - F_{i-1,j}\right) \right.\\
  &\left.
  +\frac{1}{30}\left(\tilde{F}_{i+3/2,j}-\tilde{F}_{i-3/2,j}\right)
  \right]
  \end{aligned} 
\end{equation}
where $\tilde{F}_{i+1/2,j}$ are numerical fluxes at midpoints between nodes. These midpoint-fluxes are computed by fifth order accurate WENO interpolation combined with Roe flux solver, thus makes the overall accuracy fifth order. The Mach number distribution at time $t=0.4$ obtained on unstructured and structured meshes are presented in Fig.~\ref{2dsod_comp}. Fig.~\ref{2dsod_comp_a} gives the solution on unstructured mesh using the method that we developed, while Fig.~\ref{2dsod_comp_b} is the solution obtained on structured grid of $55\times55$, which has a similar resolution to the one used in Fig.~\ref{2dsod_comp_a}. They present similar Mach number distribution, while unstructured method shows less dissipation as the red arrow in Fig.~\ref{2dsod_comp_a} indicates, where four low Mach number regions are found. Those features are smeared in Fig.~\ref{2dsod_comp_b} due to the numerical dissipation of the scheme. To further confirm this conclusion, a finer structured grid with the size of $201\times201$ is utilized with WCNS scheme. The solution is in Fig.~\ref{2dsod_comp_c}. It can be found that the abovementioned low Mach number region becomes clearer than what is found in the coarser grid, which verifies that the low-dissipation feature of the developed method is not a non-physical one. This case demonstrates that present method has the ability to capture strong discontinuities such as shock waves.

%subsection
\subsection{Shock-Vortex Interaction}
\begin{figure}[hbtp!]
  \centering
  \includegraphics[width=0.4\textwidth]{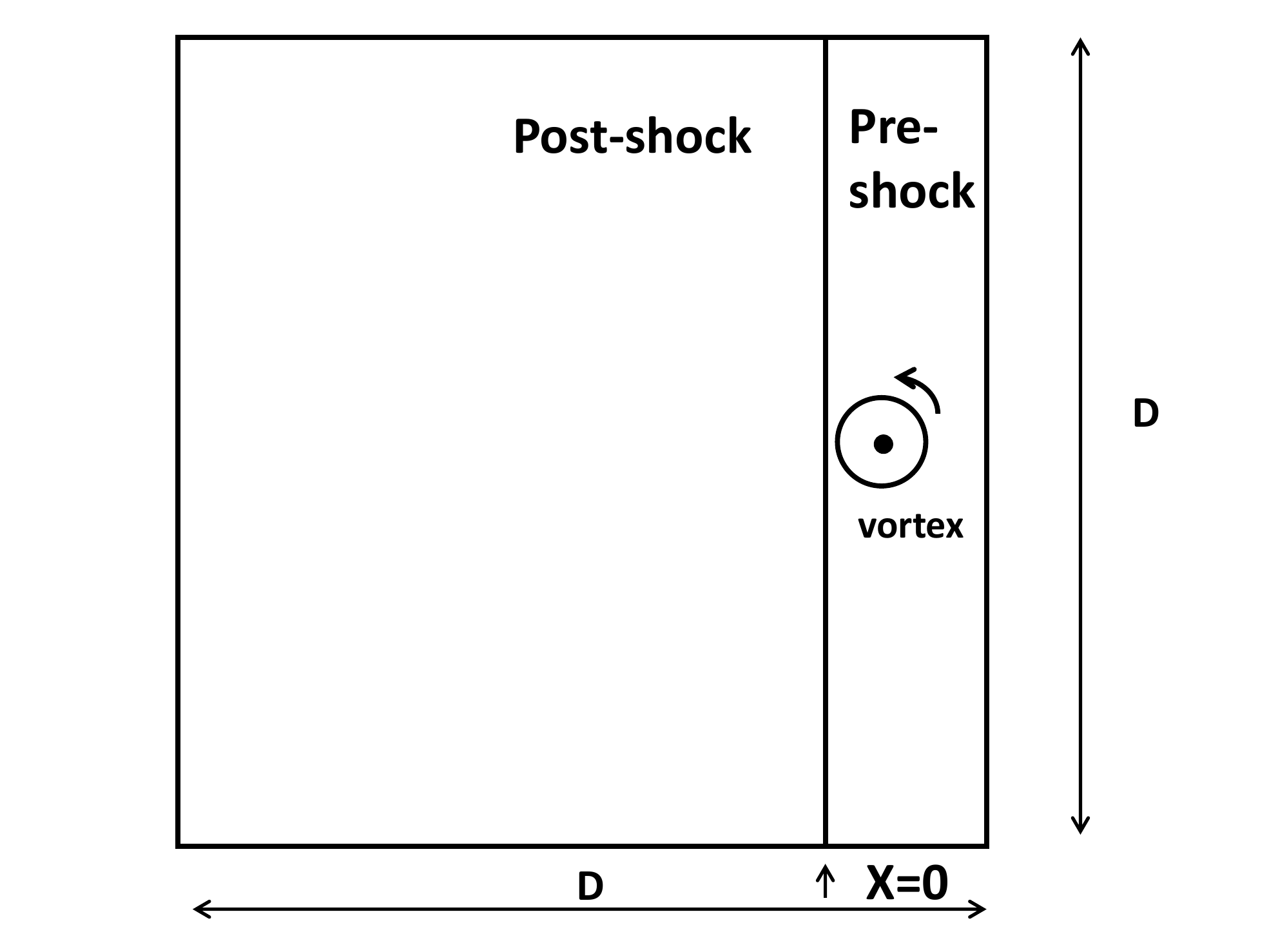}
  \caption{Shock-vortex interaction problem configuration at the initial time.}
  \label{svi_confg}
\end{figure}

\begin{figure}[htbp!]
  \centering
  \subfigure[]{
    \includegraphics[width=0.4\textwidth]{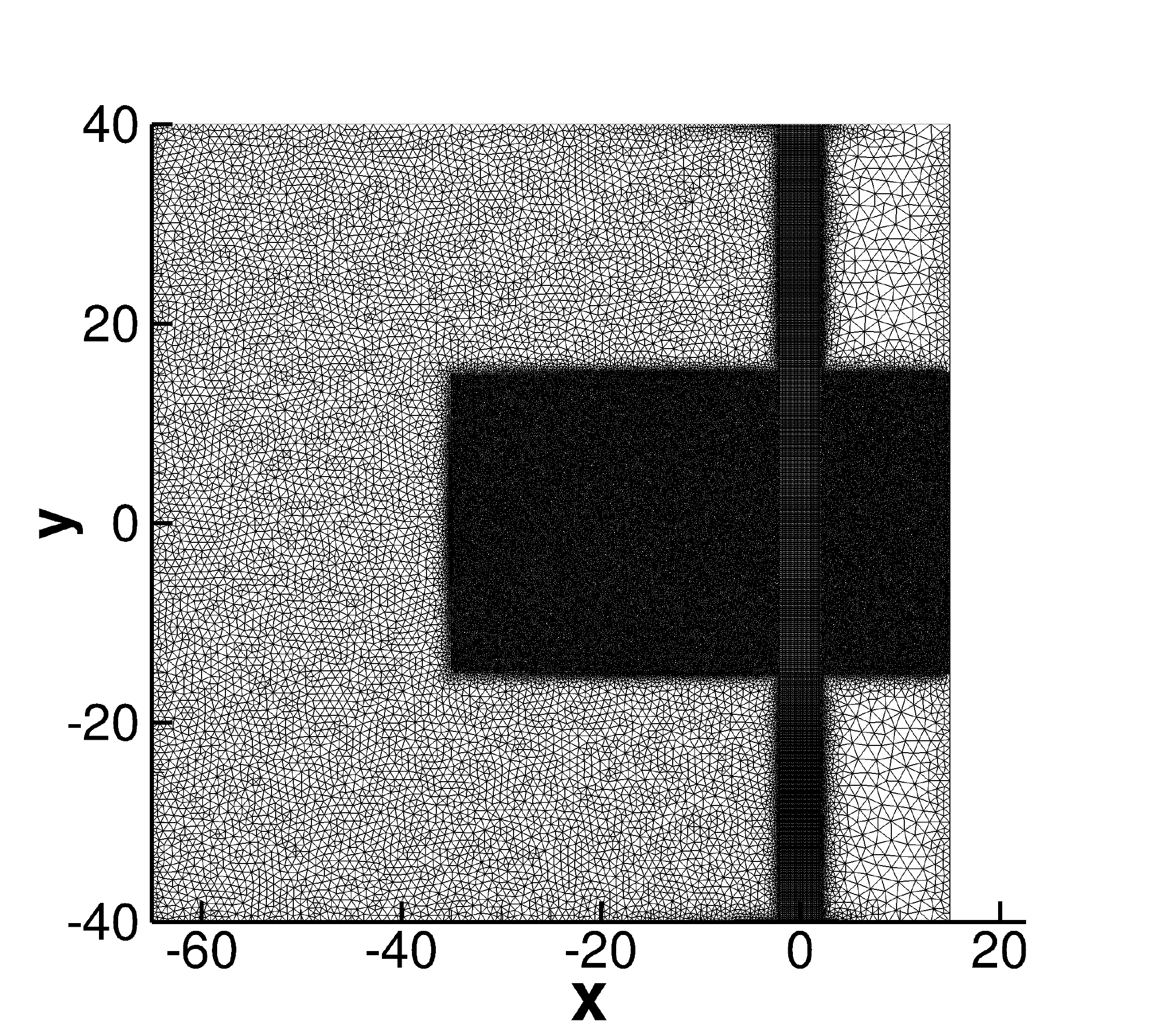}
    \centering
  }
  \subfigure[]{
    \includegraphics[width=0.4\textwidth]{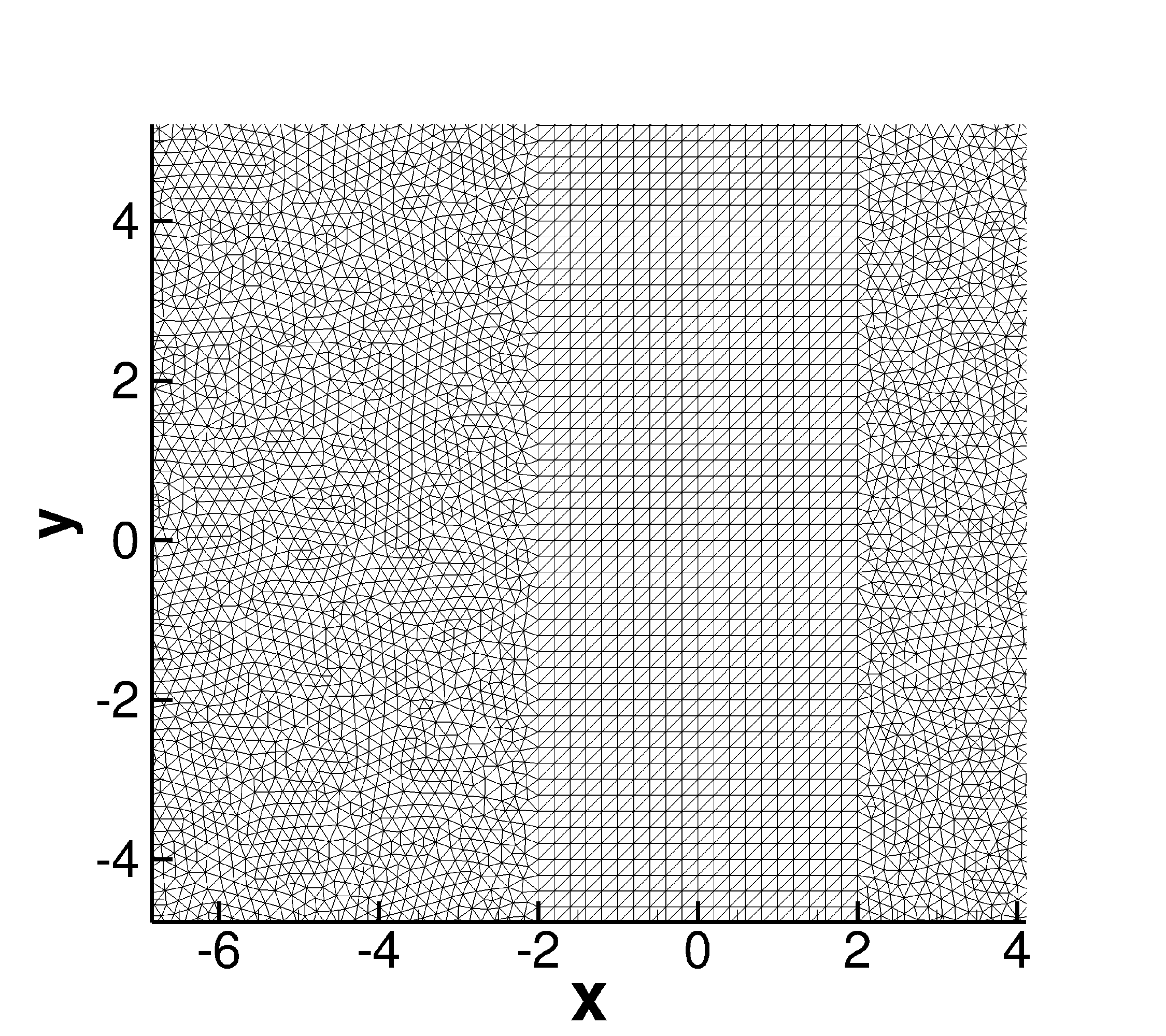}
  }
  \caption{Unstructured triangular mesh used in shock-vortex interaction problem: (a) entire mesh; (b) enlarged view of the mesh near the shock-vortex interaction region.}
  \label{svi_mesh}
\end{figure}

Shock-vortex interaction problem studied in \cite{chatterjee_multiple_2008} is numerically tested herein. This problem consists a stationary shock of Mach 1.2 and a strong isentropic vortex that convects downstream and interacts with the former. The initial configuration is shown in Fig.~\ref{svi_confg}. The computation domain size is $80\times80$ and $x\in[-65,15]$, $y\in[-40,40]$, with the core of the vortex located at $(4,0)$ initially. The stationary shock wave is at the position of $x=0$. 
Initial flow field is given by
\[
\begin{aligned}
  \rho=&\left(1-0.5(\rho-1)M_v^2e^{1-r^2}\right)^{\frac{1}{\gamma-1}}\\
  p=&\frac{1}{\gamma}\rho^\gamma\\
  (\delta u, \delta v) = &M_ve^{\frac{1-r^2}{2}}(-\bar{y},\bar{x}),\quad
  \bar{x}=x-x_c,\,\bar{y}=y-y_c
\end{aligned}
\]
where $M_v=1.0$, $(x_c,y_c)$ is the position of the vortex core, ie. (4,0). The ratio of specific heats is chosen to been $\gamma=1.4$ in this test case. Dirichlet boundary conditions are set at the left and the right boundaries, while periodic boundary conditions are applied in y direction. Unstructured mesh as in Fig.~\ref{svi_mesh} is used. The mesh is refined at the vortex convection path, interaction region and near the shock. The finest edge length in this mesh is around $\Delta x=0.08$. 

\begin{figure}[htb!]
  \centering
  \subfigure[]{
    \includegraphics[width=0.4\textwidth]{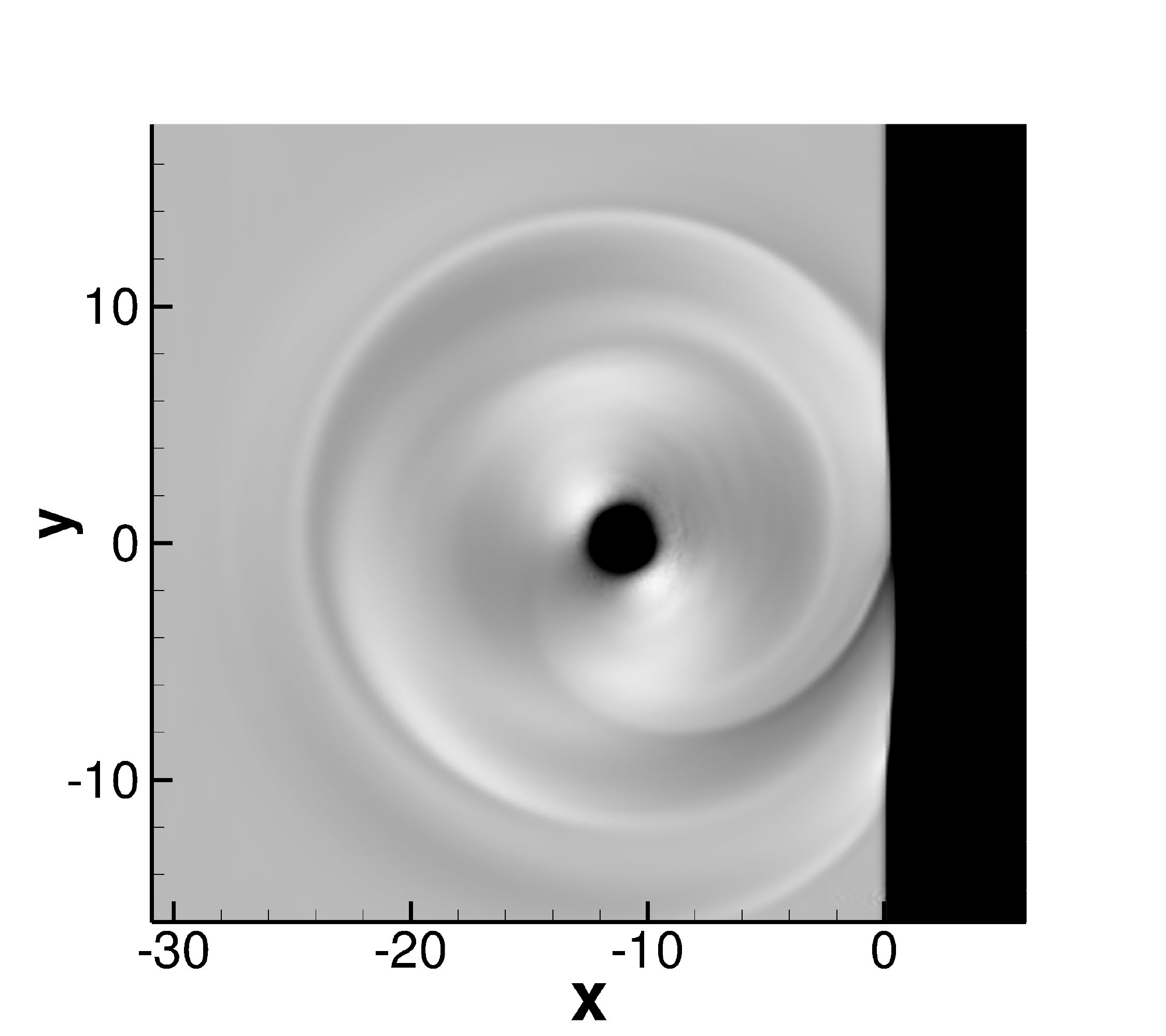}
    \label{svi_uns}
  }
  \subfigure[]{
    \includegraphics[width=0.4\textwidth]{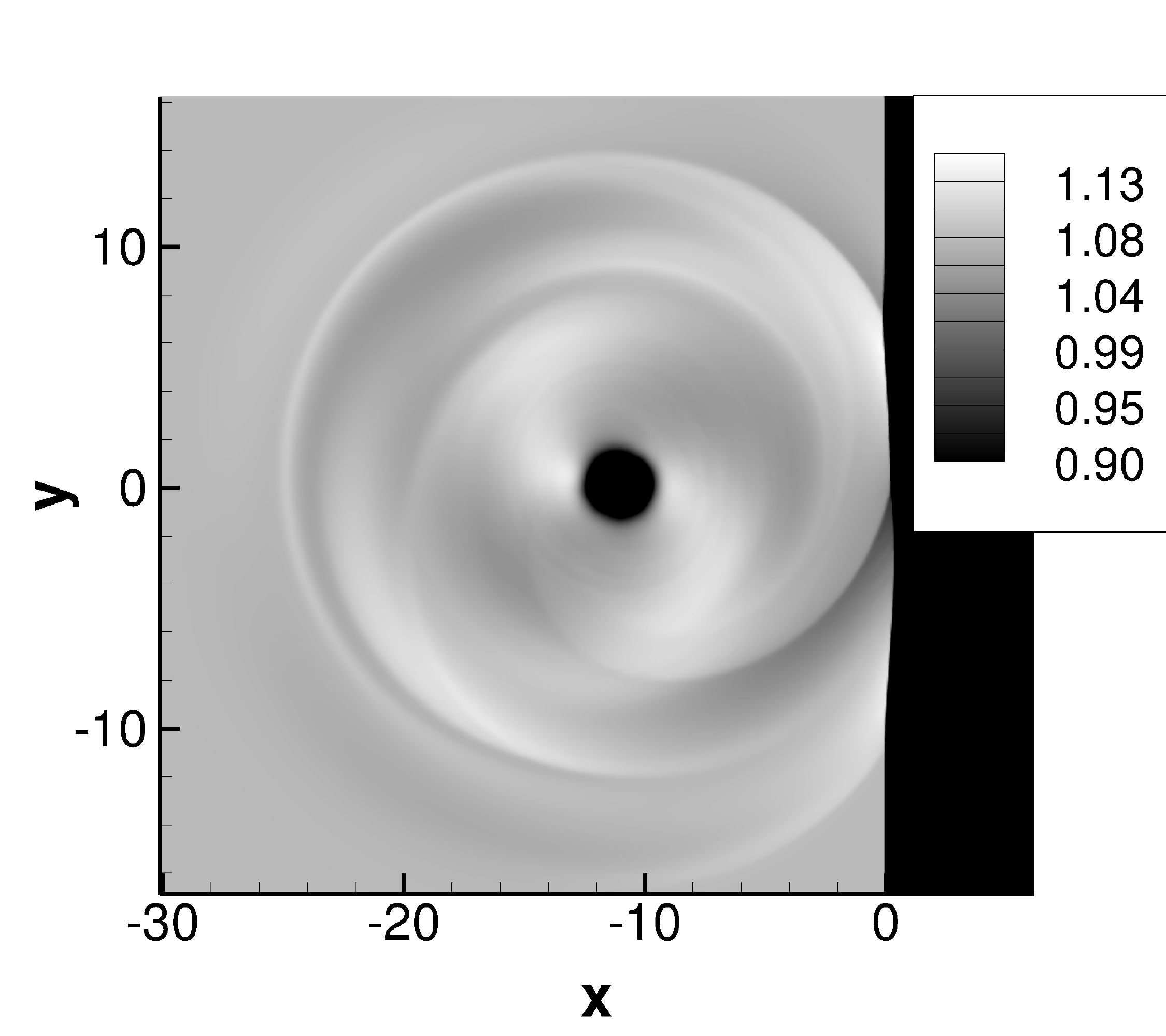}
    \label{svi_1024}
  }
  \caption{Continuous pressure contour from 0.9 to 1.15 at $t=16$ of shock-vortex interaction: (a) computed by present method; (b) computed by WCNS on structured grid.}
\end{figure}

Shock-vortex interaction at $t=16$ is shown in Fig.~\ref{svi_uns}. The shock is distorted, multiple pressure waves and reflected shocks are produced. This problem is also computed by WCNS method described in \ref{sec_2d_Sod} on a uniform structured grid for comparation. The structured grid has the resolution of $1024\times1024$ with the same case setup as the unstructured one. The computed pressure contour is shown in Fig.~\ref{svi_1024}, which gives similar interaction features.

\section{Conclusion} \label{sec4}
In this paper, we extend finite difference method to unstructured meshes. This method constructs one-dimensional stencils first to get numerical flux projection along each edge, which is similar to the way WCNS does on structured grids. It introduces the feature of highly efficient spatial discretization of structured finite difference methods, which makes this method has the potential to largely reduce the computational cost in the context of unstructured meshes. Non-uniform WENO scheme is derived to deal with non-uniform one-dimensional stencils. Based on fluxes at edges and vertices, least square method computes flux divergence and then Euler equations can be advanced readily. Owing to the use of fluxes both at edges and at vertices, the divergence stencil is quite compact.

Several numerical tests are conducted and the results show that the present method has the potential to accurately solve the flow field and to capture strong solution discontinuities. In addition, the developed method can be easily implemented in existing unstructured flow solver frames. This method shows expected accuracy on meshes from diagonalizing uniform structured grids. However, the high order performance degenerates with arbitrary unstructured meshes, which is tested and explain in detail. How to improve the interpolation accuracy will be left to the future research.

In this paper, the attention is focused on presenting a novel idea of finite difference method in the context of unstructured meshes. The method constructs one-dimensional stencils in unstructured meshes and uses least square method to get the divergence of flux vectors, which form a compute simulation method for the inviscid compressible fluid problems. However, it does not give abundant and detailed performance tests of the method. More detailed study of the efficiency, accuracy and stability of this method will be conducted in future work. Besides, numerical examples studied in the present work all consist simple boundary conditions such as periodic or Dirichlet boundary conditions. In our future work, problems in engineering level with walls or other general boundary conditions will be studied. 

\appendix
\label{appendix}
\setcounter{table}{0}
\setcounter{figure}{0}
\setcounter{equation}{0}
\renewcommand{\thetable}{A\arabic{table}}
\renewcommand{\thefigure}{A\arabic{figure}}
\renewcommand{\theequation}{A\arabic{equation}}

\section*{Appendix}
\label{sec_nu_weno}
WENO interpolation on non-uniform stencils is described herein. Points in one-dimensional stencils assembled in unstructured meshes are not equally spaced along the curve. Directly applying standard WENO interpolation 
\cite{jiang_efficient_1996} to these stencils leads to a considerably large error. This is because the standard WENO is derived from a uniform stencil. To get rid of this problem, non-uniform WENO interpolation is derived in this section. 
\begin{figure}[hbtp!]
  \centering
  \includegraphics[width=0.5\textwidth]{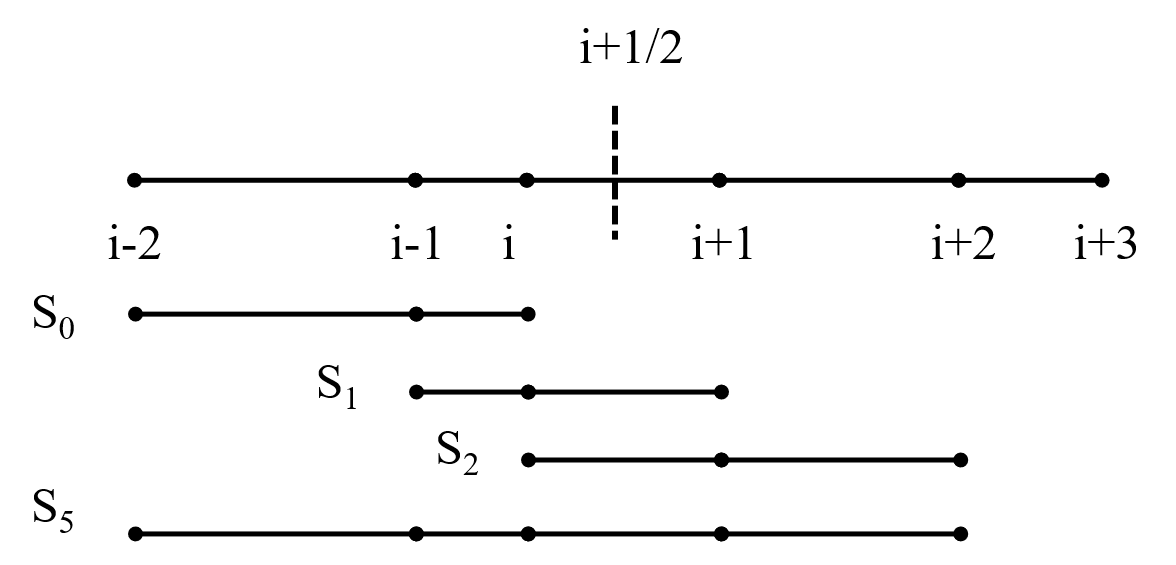}
  \caption{Stencils for non-uniform WENO to approximate value u at $x_{i+1/2}$.}
  \label{non-uniform_stencil}
\end{figure}

In order to obtain a fifth order accurate WENO interpolation, a six-point stencil is used to get the left and right states of an midpoint. Each state is nonlinearly weighted from the interpolates of three sub-stencils as shown in Fig.~\ref{non-uniform_stencil}. The stencil is basically the same as the one in \cite{jiang_efficient_1996} except that non-uniform points are used herein. For simplicity, only the left state interpolation of midpoint $x_{i+1/2}$ is presented here. The right state can be obtained by a similar way using the mirrored stencil. 
% 三个子模板得到的结果
% 5点最优线性结果
% 最优子模板加权系数
% 光滑因子定义
% 最终组合
% 
The coordinate origin of points in Fig.~\ref{non-uniform_stencil} is put at the midpoint $x_{i+1/2}$ i.e. $(x_{i}+x_{i+1})/2$, so that $x_{i+1/2}=0$. All coordinates of points are normalized by the length scale of the stencil to reduce the error introduced by numerical operation. Values at nodes like $x_i$ et al. for interpolation are written as $u_i$ et al. They are known before the upwind interpolation. These values can be primitive variables, conservative variables or other variables. Characteristic variables are used in this paper for the better stability. Each sub-stencil approximates the solution with a quadratic polynomial using three known values. These polynomials are then evaluated at the midpoint $x_{i+1/2}=0$, which give
\begin{equation}
  \begin{aligned}
    \label{u0}
    &\tilde{u}_{0,i+1/2}=\frac{a_0}{d_0}u_{i-2}+\frac{a_1}{d_0}u_{i-1}+\frac{a_2}{d_0}u_i,
  \end{aligned}
\end{equation}
where
\begin{equation*}
  \begin{aligned}
    % \centering
  &a_0=x_{i-1}x_{i}(x_i-x_{i-1}),\ a_1=-x_{i-2}x_i(x_i-x_{i-2}),\ a_2=x_{i-2}x_{i-1}(x_{i-1}-x_{i-2})\\
  &d_0=(x_{i-1}-x_{i-2})(x_i-x_{i-2})(x_i-x_{i-1}).
  \end{aligned}
\end{equation*}
\begin{equation}
  \tilde{u}_{1,i+1/2}=\frac{b_0}{d_1}u_{i-1}+\frac{b_1}{d_1}u_{i}+\frac{b_2}{d_1}u_{i+1},
\end{equation}
where
\begin{equation*}
  \begin{aligned}
    \label{u2}
    &b_0=x_{i}x_{i+1}(x_{i+1}-x_{i}),\ b_1=-x_{i-1}x_{i+1}(x_{i+1}-x_{i-1}),\ b_2=x_{i}x_{i-1}(x_{i}-x_{i-1})\\
    &d_1=(x_{i}-x_{i-1})(x_{i+1}-x_{i-1})(x_{i+1}-x_{i}).
  \end{aligned}
\end{equation*}
\begin{equation}
  \tilde{u}_{2,i+1/2}=\frac{c_0}{d_2}u_{i}+\frac{c_1}{d_2}u_{i+1}+\frac{c_2}{d_2}u_{i+2},
\end{equation}
where
\begin{equation*}
  \begin{aligned}
    &c_0=x_{i+1}x_{i+2}(x_{i+2}-x_{i+1}),\ c_1=-x_{i}x_{i+2}(x_{i+2}-x_{i}),\ c_2=x_{i+1}x_{i}(x_{i+1}-x_{i})\\
    &d_2=(x_{i+1}-x_{i})(x_{i+2}-x_{i})(x_{i+2}-x_{i+1}).
  \end{aligned}
\end{equation*}
It is noticed that the interpolated value at midpoint is the weighted summation of values at nodes, which has the same form as in the uniform WENO case, except that the coefficients of values now are calculated from coordinates of points rather than constant numbers.
The large stencil $S_5$ of five points in Fig.~\ref{non-uniform_stencil} utilizes a fourth-degree polynomial to approximate the solution. Five coefficients of the polynomial are determined by interpolation using values at five points. The polynomial gives the value at $x_{i+1/2}$ as
\begin{equation}
  \label{S5}
  \tilde{u}_{5,i+1/2} = \frac{l_0}{d_5}u_0 + \frac{l_1}{d_5}u_1 + \frac{l_2}{d_5}u_2 + \frac{l_3}{d_5}u_3 + \frac{l_4}{d_5}u_4,
\end{equation}
where
\begin{equation*}
  \begin{aligned}
    l_0 = &x_{i-1}x_{i}x_{i+1}x_{i+2}\\
    &\cdot(x_{i}-x_{i-1})(x_{i+1}-x_{i-1})(x_{i+1}-x_{i})(x_{i+2}-x_{i-1})(x_{i+2}-x_{i})(x_{i+2}-x_{i+1}),\\
    l_1 =& -x_{i-2}x_{i}x_{i+1}x_{i+2}(x_{i}-x_{i-2})\\
    &\cdot(x_{i+1}-x_{i-2})(x_{i+1}-x_{i})(x_{i+2}-x_{i-2})(x_{i+2}-x_{i})(x_{i+2}-x_{i+1}),  \\
    l_2 = &x_{i-2}x_{i-1}x_{i+1}x_{i+2}\\
    &\cdot(x_{i-1}-x_{i-2})(x_{i+1}-x_{i-2})(x_{i+1}-x_{i-1})(x_{i+2}-x_{i-2})(x_{i+2}-x_{i-1})(x_{i+2}-x_{i+1}),\\
    l_3 = &-x_{i-2}x_{i-1}x_{i}x_{i+2}\\
    &\cdot(x_{i-1}-x_{i-2})(x_{i}-x_{i-2})(x_{i}-x_{i-1})(x_{i+2}-x_{i-2})(x_{i+2}-x_{i-1})(x_{i+2}-x_{i}), \\ 
    l_4 = &x_{i-2}x_{i-1}x_{i}x_{i+1}\\
    &\cdot(x_{i-1}-x_{i-2})(x_{i}-x_{i-2})(x_{i}-x_{i-1})(x_{i+1}-x_{i-2})(x_{i+1}-x_{i-1})(x_{i+1}-x_{i}), \\
    d_5=&(x_{i-1}-x_{i-2})(x_{i}-x_{i-2})(x_{i}-x_{i-2})(x_{i+1}-x_{i-2})(x_{i+1}-x_{i-1})\\
    &\cdot(x_{i+1}-x_{i})(x_{i+2}-x_{i-2})(x_{i+2}-x_{i-1})(x_{i+2}-x_{i})(x_{i+2}-x_{i+1}).
  \end{aligned}
\end{equation*}
It is noticed that coefficients in Eq.~\eqref{S5} have a certain regularity and can be simply expressed as
\begin{equation}
  \begin{aligned}
  l_k&=(-1)^k\prod_{p=i-2,\atop p\neq i-2+k}^{i+2}x_p \cdot \prod_{m,n=i-2,\atop m<n,\ m,n\neq i-2+k}^{i+2}(x_n-x_m),\quad k=0,1,\dots4\\
  d_5&=\prod_{m,n=i-2,\atop m<n}^{i+2}(x_j-x_i),
\end{aligned}
\end{equation}
which also apply to Eq.~\eqref{u0}~Eq.\eqref{u2} with minimal change to the lower and upper limits of $i$ and $j$.

In order to get higher order accuracy, three sub-stencils $S_0$, $S_1$, $S_2$ are combined together to get the value at $x_{i+1/2}$, given as
\begin{equation}
  \label{u_half}
  \tilde{u}_{i+1/2}=\sum_{k=0}^{2}C_k\tilde{u}_{k,i+1/2},
\end{equation}
where $C_k$ are the linear weights which can be obtained as follows.  $\tilde{u}_{i+1/2}$ in Eq.~\ref{u_half} can be expanded to the weighted sum of $u_i$ etc. As $u_i$ etc. are arbitrary, the constant coefficients of them should be the same as in Eq.~\ref{S5}. The optimum weights read
\begin{equation}
  \begin{aligned} 
    \label{C_K}
  &C_0=\frac{x_{i+1}x_{i+2}}{(x_{i+1}-x_{i-2})(x_{i+2}-x_{i-2})},\\
  &C_1=\frac{-x_{i-2}(-x_{i-2}-x_{i-1}+x_{i+1}+x_{i+2})x_{i+2}}{(x_{i+1}-x_{i-2})(x_{i+2}-x_{i-2})(x_{i+2}-x_{i-1})},\\
  &C_2=\frac{x_{i-2}x_{i-1}}{(x_{i+2}-x_{i-2})(x_{i+2}-x_{i-1})}.
\end{aligned}
\end{equation}
Note that $x_{i+1/2}=(x_{i}+x_{i+1})/2=0$ is used during the above derivation. It can be verified readily that Eq.~\eqref{C_K} satisfies
\begin{equation}
  C_0+C_1+C_2=1
\end{equation}
The scheme we derived so far is a linear upwind one. In order to capture strong discontinuities, nonlinearity should be taken into consideration by modifying optimal weights in Eq.~\eqref{C_K} using smooth indicators.

The smooth indicator defined herein is of the same form in \cite{jiang_efficient_1996} by
\begin{equation}
\beta_k=\sum_{l=1}^{2}\int_{x_i}^{x_{i+1}} \Delta x^{2l-1}\left(\frac{\partial^l\tilde{u}_k(x)}{\partial x^l}\right)^2\,dx,\ k=0,1,2.
\end{equation}
The derivation of $\beta_k$ is straightforward by using $\tilde{u}_{k,i}$ at hand. The nonlinear interpolation of $u$ at midpoint is given by
\begin{equation}
  \tilde{u}_{i+1/2}=\sum_{k=0}^2\omega_k \tilde{u}_{k,i+1/2},
\end{equation}
where $\omega_k$ are nonlinear weights formulated by
\begin{equation}
  \omega_k=\frac{\alpha_k}{\sum_{i=0}^2\alpha_i},\ \alpha_k=\frac{C_k}{(\beta_k+\epsilon)^p},\ k=0,1,2,
\end{equation}
where $C_k$ are linear optimal weights in Eq.~\eqref{C_K} and $p$ is a positive integer, which is chosen to be one in this paper. $\epsilon=1.0e^{-40}$ is a small number used in case that the equation is divided by zero. The nonlinear weights are designed that in smooth regions of the solution, weights are almost the same as linear weights, which produces a scheme of fifth order accuracy. However, if a strong discontinuity lies in an interpolation stencil, smooth indicator would become extremely large that the weight of which would be near zero. Thus, the overall accuracy become third-order accurate as only a part of sub-stencils are being used for interpolation.
%%%% Acknowledgments %%%%%%%%
\section*{Acknowledgments}
The authors acknowledge the support from the National Natural Science Foundation of China (Grant No. 91952203).

\bibliographystyle{abbrv}
\bibliography{main}

\begin{thebibliography}{10}

\bibitem{barth_higher_1990}
T.~Barth and P.~Frederickson.
\newblock Higher order solution of the {Euler} equations on unstructured grids
  using quadratic reconstruction.
\newblock In {\em 28th {Aerospace} {Sciences} {Meeting}}, Reno,NV,U.S.A., Jan.
  1990. American Institute of Aeronautics and Astronautics.

\bibitem{borges_improved_2008}
R.~Borges, M.~Carmona, B.~Costa, and W.~S. Don.
\newblock An improved weighted essentially non-oscillatory scheme for
  hyperbolic conservation laws.
\newblock {\em Journal of Computational Physics}, 227(6):3191--3211, Mar. 2008.

\bibitem{chatterjee_multiple_2008}
A.~Chatterjee and S.~Vijayaraj.
\newblock Multiple {Sound} {Generation} in {Interaction} of {Shock} {Wave} with
  {Strong} {Vortex}.
\newblock {\em AIAA Journal}, 46(10):2558--2567, Oct. 2008.

\bibitem{cockburn1989tvb}
B.~Cockburn and C.-W. Shu.
\newblock Tvb runge-kutta local projection discontinuous galerkin finite
  element method for conservation laws. ii. general framework.
\newblock {\em Mathematics of computation}, 52(186):411--435, 1989.

\bibitem{cockburn2001runge}
B.~Cockburn and C.-W. Shu.
\newblock Runge--kutta discontinuous galerkin methods for convection-dominated
  problems.
\newblock {\em Journal of scientific computing}, 16(3):173--261, 2001.

\bibitem{deng_developing_2000}
X.~Deng and H.~Zhang.
\newblock Developing {High}-{Order} {Weighted} {Compact} {Nonlinear} {Schemes}.
\newblock {\em Journal of Computational Physics}, 165(1):22--44, Nov. 2000.

\bibitem{henrick_mapped_2005}
A.~K. Henrick, T.~D. Aslam, and J.~M. Powers.
\newblock Mapped weighted essentially non-oscillatory schemes: {Achieving}
  optimal order near critical points.
\newblock {\em Journal of Computational Physics}, 207(2):542--567, Aug. 2005.

\bibitem{hu_weighted_1999}
C.~Hu and C.-W. Shu.
\newblock Weighted {Essentially} {Non}-oscillatory {Schemes} on {Triangular}
  {Meshes}.
\newblock {\em Journal of Computational Physics}, 150(1):97--127, Mar. 1999.

\bibitem{hu_adaptive_2010}
X.~Hu, Q.~Wang, and N.~Adams.
\newblock An adaptive central-upwind weighted essentially non-oscillatory
  scheme.
\newblock {\em Journal of Computational Physics}, 229(23):8952--8965, Nov.
  2010.

\bibitem{huynh_flux_2007}
H.~T. Huynh.
\newblock A {Flux} {Reconstruction} {Approach} to {High}-{Order} {Schemes}
  {Including} {Discontinuous} {Galerkin} {Methods}.
\newblock In {\em 18th {AIAA} {Computational} {Fluid} {Dynamics} {Conference}},
  Miami, Florida, June 2007. American Institute of Aeronautics and
  Astronautics.

\bibitem{jiang_efficient_1996}
G.-S. Jiang and C.-W. Shu.
\newblock Efficient {Implementation} of {Weighted} {ENO} {Schemes}.
\newblock {\em Journal of Computational Physics}, 126(1):202--228, June 1996.

\bibitem{lacaze2009large}
G.~Lacaze, B.~Cuenot, T.~Poinsot, and M.~Oschwald.
\newblock Large eddy simulation of laser ignition and compressible reacting
  flow in a rocket-like configuration.
\newblock {\em Combustion and Flame}, 156(6):1166--1180, 2009.

\bibitem{li_high_2020}
X.-L. Li and Y.-X. Ren.
\newblock High order compact generalized finite difference methods for solving
  inviscid compressible flows.
\newblock 82(1):18.

\bibitem{liu2006spectral}
Y.~Liu, M.~Vinokur, and Z.~J. Wang.
\newblock Spectral difference method for unstructured grids i: basic
  formulation.
\newblock {\em Journal of Computational Physics}, 216(2):780--801, 2006.

\bibitem{reed1973triangular}
W.~H. Reed and T.~Hill.
\newblock Triangular mesh methods for the neutron transport equation.
\newblock Technical report, Los Alamos Scientific Lab., N. Mex.(USA), 1973.

\bibitem{shu_efficient_1988}
C.-W. Shu and S.~Osher.
\newblock Efficient implementation of essentially non-oscillatory
  shock-capturing schemes.
\newblock {\em Journal of Computational Physics}, 77(2):439--471, Aug. 1988.

\bibitem{sod1978survey}
G.~A. Sod.
\newblock A survey of several finite difference methods for systems of
  nonlinear hyperbolic conservation laws.
\newblock {\em Journal of computational physics}, 27(1):1--31, 1978.

\bibitem{su2020numerical}
H.~Su, J.~Cai, K.~Qu, and S.~Pan.
\newblock Numerical simulations of inert and reactive highly underexpanded
  jets.
\newblock {\em Physics of Fluids}, 32(3):036104, 2020.

\bibitem{wang_compact_2016}
Q.~Wang, Y.-X. Ren, and W.~Li.
\newblock Compact high order finite volume method on unstructured grids {II}:
  {Extension} to two-dimensional {Euler} equations.
\newblock {\em Journal of Computational Physics}, 314:883--908, June 2016.

\bibitem{wang2007spectral}
Z.~J. Wang, Y.~Liu, G.~May, and A.~Jameson.
\newblock Spectral difference method for unstructured grids ii: extension to
  the euler equations.
\newblock {\em Journal of Scientific Computing}, 32(1):45--71, 2007.

\bibitem{witherden_pyfr_2014}
F.~Witherden, A.~Farrington, and P.~Vincent.
\newblock {PyFR}: {An} open source framework for solving advection–diffusion
  type problems on streaming architectures using the flux reconstruction
  approach.
\newblock {\em Computer Physics Communications}, 185(11):3028--3040, Nov. 2014.

\bibitem{wong_high-order_2017}
M.~L. Wong and S.~K. Lele.
\newblock High-order localized dissipation weighted compact nonlinear scheme
  for shock- and interface-capturing in compressible flows.
\newblock 339:179--209.

\bibitem{yee_low-dissipative_1999}
H.~Yee, N.~Sandham, and M.~Djomehri.
\newblock Low-{Dissipative} {High}-{Order} {Shock}-{Capturing} {Methods}
  {Using} {Characteristic}-{Based} {Filters}.
\newblock {\em Journal of Computational Physics}, 150(1):199--238, Mar. 1999.

\end{thebibliography}

\end{document}